\documentclass[12pt,manuscript]{aastex}

\def\ngc#1{\hbox{NGC$\,$#1}}

\def\etal{{et~al.}}
\def\ie{{\it i.e.}}
\def\eg{{\it e.g.}}
\def\cf{{\it cf.}}
\def\rnum#1{{\uppercase\expandafter{\romannumeral#1}}}

\def\hr{${}^{\rm h}$}
\def\mn{${}^{\rm m}$}

\def\Sc{${}^{\rm s}$\llap{.}}

\def\deg{${}^\circ$}
\def\min{${}^{\prime}$}
\def\sec{${}^{\prime\prime}$}
\def\ltsim{ \,{}^<_\sim\, }
\def\gtsim{ \,{}^>_\sim\, }

\def\umi{\hbox{\it U--I\/}}
\def\vmi{\hbox{\it V--I\/}}
\def\vmr{\hbox{\it V--R\/}}
\def\rmi{\hbox{\it R--I\/}}
\def\bmv{\hbox{\it B--V\/}}
\def\umb{\hbox{\it U--B\/}}
\def\bmi{\hbox{\it B--I\/}}

\def\today{\number\year\space \ifcase\month\or
  January\or February\or March\or April\or May\or June\or
  July\or August\or September\or October\or November\or December\fi
  \space\number\day}
\def\now{\number\year\space \ifcase\month\or
  January\or February\or March\or April\or May\or June\or
  July\or August\or September\or October\or November\or December\fi
  \space\number\day .\number\time}

\begin{document}

\received{}
\revised{}
\accepted{}
\ccc{}
\cpright{}{}

\slugcomment{}
\shorttitle{\ngc{2419}}
\shortauthors{P.~B.~Stetson}

\title{Homogeneous Photometry IV:  On the Standard Sequence in the Globular
Cluster \ngc{2419}\footnote{Based in part on observations obtained at the 3.5m
and 0.9m WIYN Telescopes. The WIYN Observatory is a joint facility of the
University of Wisconsin-Madison, Indiana University, Yale University, and the
National Optical Astronomy Observatory (NOAO).}}

\author{Peter B. Stetson\altaffilmark{2,}\altaffilmark{3}}
\affil{Dominion Astrophysical Observatory, Herzberg Institute of Astrophysics,\\
National Research Council, 5071 West Saanich Road, Victoria, BC V9E 2E7, Canada;
Peter.Stetson@nrc.gc.ca}

\altaffiltext{2}{Guest Investigator of the UK Astronomy Data Centre.}

\altaffiltext{3}{Guest User, Canadian Astronomy Data Centre, which is operated
by the Herzberg Institute of Astrophysics, National Research Council of Canada.}

\received{}
\revised{}
\accepted{}

\begin{abstract}

A new analysis of CCD-based {\it BVRI\/} broad-band photometry for the globular
cluster \ngc{2419}\ is presented, based on 340 CCD images either donated by
colleagues or retrieved from public archives.  The calibrated results have been
made available through my Web site.  I compare the results of my analysis with
those of an independent analysis of a subset of these data by Saha \etal\ (2005,
PASP, 117, 37), who have found a color-dependent discrepancy of up to
0.05$\,$mag between their $I$-band photometry and mine for stars in this
cluster.  I conclude that a major part of the discrepancy appears to be
associated with small (a few hundredths of a second) shutter-timing errors
in the MiniMos camera on the WIYN 3.5-m telescope.  Smaller contributions to
the anomaly likely come from (a)~a color-scale error with a maximum amplitude
of $\sim\pm0.02\,$mag in my secondary standard list as of September 2004;
and (b)~statistical effects arising from the previous study's use of a
relatively small number of standard-star observations to determine a
comparatively large number of fitting parameters in the photometric
transformations.  

\end{abstract}


\section{INTRODUCTION}

The calibration of broad-band photometric measurements to a uniform system is an
ongoing challenge in observational astronomy.  The photometric systems in
general use are tied to networks of standard stars whose relative brightnesses
in different bandpasses have been repeatedly measured in the course of dedicated
programs with carefully monitored equipment.  Other observers can use their own
equipment to measure the brightness ratios between as many as possible of these
stars and the astronomical targets of their specific research programs. 
Obtaining precision in astronomical photometry is largely a matter of collecting
sufficient photons under excellent observing conditions on numerous occasions
over a long time, so that transient variations such as scintillation and
erratic changes in the transparency of the terrestrial atmosphere eventually
average out.  

Obtaining accuracy in astronomical photometry is often a more demanding task.
It requires the best possible match between the observer's instrumental
bandpasses and those employed by the establishers of the photometric standard
system being used.  In addition, the task requires that observations be made of
numerous photometric standard stars under conditions that are as identical as
possible to those under which the target observations are made.  Enough
standard-star observations must be made each night to allow the detection and
{\it ex post facto\/} removal of systematic trends due to residual bandpass
mismatch, the quasi-static properties of the atmosphere, and any other drifts or
nonlinearities in the equipment.  The resulting photometry can be no more
accurate than the level to which these systematic errors can be identified,
measured, and calibrated away.

The problem of establishing standard photometric systems has not been solved for
all time.  The last few decades have seen the proliferation of telescopes with
4m, 8m, and 10m apertures, and the quantum efficiency of detectors is now close
to 100\%.  To achieve photometric {\it accuracy\/} to a level similar to the
{\it precision\/} that is possible with modern equipment, it is necessary to
have available many standard stars with as broad as possible a range of
properties (primarily temperature, secondarily surface gravity and chemical
abundances).  It is desirable for these stars to be similar in apparent
brightness to the targets of modern research programs to minimize residual
instrumental nonlinearity, but there is a competing temptation to define
and employ relatively bright standard stars to minimize integration times. 
Standard stars should be distributed as widely as possible on the sky to
minimize the time needed to slew the telescope and also to probe the same
sightlines through the terrestrial atmosphere as the program observations, but
they should also be densely packed on the sky to permit the simultaneous
observation of standards with a wide range of intrinsic properties.  

Arlo Landolt (1973, 1983, 1992) has devoted the greater part of his research
career to establishing a network of standard stars in the Johnson {\it UBV\/}
system and, more recently, the {\it RI\/} system defined by Kron, White \&
Gascoigne (1953) and subsequently extended and modified by Cousins (1976). 
Landolt refers to this as the ``Johnson-Kron-Cousins system,'' although some
others call it the ``Landolt system.''  Landolt's standard stars are
concentrated near the celestial equator so that precisely the same photometric
system is available in both geographic hemispheres, and standards are defined in
every hour of right ascension so that some are accessible both on and off the
meridian at any time of night or year.  In every third hour of right ascension a
particularly large number of standards is concentrated in a small area of sky to
simplify the efficient calibration of all parts of wide-area detectors.

Landolt's standard-system magnitudes have been established by photomultiplier
measurements, a technique which can yield an accuracy often superior to that
possible with a typical CCD camera.  First, nonlinearity in a pulse-counting
system is vanishingly small for faint targets, and still small but readily
calibrated out for relatively bright targets.  Second, flat-fielding and readout
noise are generally non-issues with photomultipliers, while the difficulty of
proper flat-fielding is often a limiting factor in CCD photometry.

Photoelectric photometry has the drawback that measuring apertures much larger
than the full-width at half-maximum of a point source must be employed to
contain the worst seeing conditions and extremes of tracking wander that may
occur during the integration.  This leads to the possibility of a systematic
error in the average photometry of any given star due to the presence of an
unseen fainter companion within the measurement aperture.  More importantly, the
technique limits the faintess of the stars that can be reliably measured, as the
photon statistics inherent in subtracting a separate ``sky'' measurement from a
``target plus sky'' measurement quickly dominate the error budget when the
target flux is small compared to the diffuse sky flux in a large measuring
aperture.  The relatively low quantum efficiency of the typical photomultiplier
compared to the best CCDs and the reluctance of many TACs to take
large-telescope time away from ``science'' for ``calibration'' are further
factors limiting the faintness of photomultiplier-defined photometric 
standards.  For these reasons, fundamental photometric standard
stars suitable for calibrating modern detectors on large telescopes or on a
medium-sized diffraction-limited telescope in space are not numerous.  

Over the years, a number of individuals and groups have endeavored to provide
faint photometric sequences for the modern age of large telescopes and
electronic detectors.  Many observers, for their own purposes, define local
standard sequences near astronomical objects of interest by repeatedly comparing
selected stars to the standards of Landolt or others, and they often tabulate
these standard sequences in their publications in hopes that they will help to
place subsequent investigations by others on a common magnitude system, at least
for that field.  Examples are the photometry in the open cluster M67 by Schild
(1983) and by Joner \& Taylor (1990), who encouraged readers to employ
their results for the cluster stars as reference standards for other research
programs.  In the course of an investigation of Cepheid variables in the nearby
Sc galaxy \ngc{300}, Graham (1981) established a faint photometric sequence in a
field near the galaxy.  Walker \& Suntzeff (1990, 1991) have similarly defined
and subsequently refined a standard sequence in the field of SN1987A.  Other
examples abound.  

However, there have also been attempts to augment the number of different
standard fields suitable for modern imaging equipment as a general service to
the community, quite apart from the scientific interest of any particular
targets.  Among these are the sequences produced by  the ``KPNO Video Camera/CCD
Standards Consortium'' (Christian \etal\ 1985).  The Consortium adopted the wise
policy of establishing standard sequences in star clusters, where an optimum
surface density of stars can be achieved simply by placing the field an
appropriate distance from the center of the cluster.  Christian \etal\ chose six
targets---two open clusters and four globular clusters---spaced at intervals of
roughly four to six hours of right ascension.  The 53 Consortium standard stars
extend as faint as $V\sim19$.  In comparison, if one considers only the Landolt
standards with at least five independent observations in {\it UBVRI\/} (thus not
counting stars with {\it UBV\/} from his 1973 paper), there are 283 such stars
and the faintest has $V=17.8$.  However, the 95${}^{th}$ percentile (269/283) of
the Landolt standards is at $V=15.1$, forty times brighter than the faintest
Consortium standard.  Unfortunately, the precision of the Consortium photometry
does not rival that of Landolt:  the root-mean-square (r.m.s.) scatter of their
observations was typically 0.02$\,$mag per measurement, and most stars had two
to four measurements apiece, yielding a typical final precision $\gtsim
0.01\,$mag for each tabulated magnitude.  Among the 283 Landolt standards with
at least five observations, the median number of observations per star in {\it
RI\/} is 20 (the median number is greater than this in {\it UBV\/} when the 1973
data are included) and the median standard error of a mean magnitude is
0.0030--0.0036$\,$mag, depending upon the filter.  Three of the six Consortium
fields lie north of +39\deg\ declination, rendering them inaccessible to the
major southern observatories, and the other three lie between +9\deg\ and
+18\deg, making them only modestly useful in the South.  

There are other {\it UBVRI\/} standard lists in the literature. For instance,
Graham (1982) published photomultiplier-based Johnson {\it UBV\/} and
Kron-Cousins {\it RI\/} photometry for 102 stars in nine E-Region fields at
--45\deg\ declination.  These have been shown to be on the Landolt system to
within an accuracy of a couple of mmag (Taylor \& Joner 1996) in at least the
{\it VRI\/} filters, which is comparable to the precision claimed for the best
of the published indices.  Like Landolt's standards these stars are
comparatively bright:  among the 64 Graham stars with five or more observations,
the 95${}^{th}$ percentile (61/64) has $V = 14.7$.  Stobie, Sagar \& Gilmore
(1985) provided CCD-based Johnson {\it BV\/} and Gunn $i$ photometry for some
512 stars with $12 < V < 22$ in seven fields between declinations --60\deg\ and
+10\deg.  They do not provide quantitative standard errors for their derived
photometric indices, but they did estimate that in a single 30-minute exposure
their root-mean-square measuring errors typically ranged from
0.02$\,$mag per observation for $I < 18$ to 0.17$\,$mag per observation for
$20 < I < 21$.  However, each field was observed four times in each filter with
different exposure times:  the authors give, as an example, 1~minute, 3~minutes,
10~minutes, and 30~minutes, but imply that these exposure times were not
strictly adhered to for all fields.  Furthermore, at least one field was
observed on two occasions.  Therefore, the actual precision of their tabulated
results is hard to evaluate.  As another example, Odewahn, Bryja \& Humphreys
(1992) obtained additional {\it BVR\/} CCD photometry for three of the KPNO
Consortium fields both to increase the number of available standard stars and to
improve the precision of the published indices.  

As of February 28, 2005, a custom query of the NASA ADS Astronomy/Planetary
database for the title words ``standard,'' ``star,'' and ``photometry'' combined
with the ``and'' option yielded 103 abstracts, so I will not attempt a thorough,
fair, and even-handed survey of the subject.  The studies I have mentioned are
probably representative or, perhaps, rather better than average.  

During the pursuit of other research goals over the past two decades, I have
maintained a parallel program of acquiring---by personally obtaining them at the
telescope, by soliciting them from colleagues, or by requesting them from public
archives---CCD images suitable for the eventual establishment of faint
photometric sequences that would be accessible to the most sensitive broad-band
imaging cameras available now or for the foreseeable future.  With these I hope
that other researchers and I will have the resources to enable us to calibrate
our measured instrumental magnitudes to a common photometric system with an
accuracy that is limited more by the equipment and observing techniques than by
the lack of suitably defined reference standards.  This would enable
observations of different astronomical targets obtained on different occasions
and with different equipment to be intercompared and interpreted with a degree
of subtlety and confidence not currently possible.  The first three papers in an
ongoing series on the subject of ``Homogeneous Photometry'' have already been
published (Stetson, Hesser \& Smecker-Hane 1998; Stetson 2000; Stetson, Bruntt
\& Grundahl 2003; hereinafter HP-I, HP-II, and HP-III); this paper is the
fourth.  HP-II, in particular, goes into the philosophy and methodology of this
work in greater depth than I will do here.  

This network of standard stars both grows and evolves with time, as new
observing runs are ingested into the system:  new standard fields are added,
more stars achieve the requisite number of observations and level of precision
to satisfy the acceptance criteria for inclusion in the standard list, and
occasionally stars are lost as new observations suggest that they might be
intrinsically variable.  It may seem surprising, but the Landolt system that I
am trying to match is itself something of a moving target, because a new
observing run sometimes includes a Landolt standard that had not previously had
any observations in my database.  This slightly alters the ``Landolt system''
that I am trying to calibrate to, since even in Landolt's data table each entry
has some unknown random error with respect to the mean system.  As new
observations arrive for some Landolt standards but not for others, the different
stars' relative influence changes, which also causes some minor migration of the
effective mean photometric system.  Of course, all these uncertainties diminish
as the number of different stars and the number of repeat observations included
in the body of data both increase.  The instantaneous state of my standard-star
list is available to the community via the World-Wide
Web\footnote{http://cadcwww.hia.nrc.ca/cadcbin/wdb/astrocat/stetson/query/}. 
Before each new data release, however, I have tried to ensure that my
photometric indices, on average, are on the Landolt system to within a very fine
tolerance.  This has been described in HP-II.  The September 28, 2004 version of
this catalog contained 36,950 stars suitable for use as standards, defined as
stars with at least five observations on photometric occasions and standard
errors of the mean $< 0.020\,$mag in at least two of the {\it BVRI\/} filters,
and no evidence of intrinsic variability in excess of 0.050$\,$mag,
r.m.s.  Of these, 1,021 stars are fainter than $V = 21.00$.  

A recent paper by Saha \etal\ (2005) also pursues the goal of establishing
reliable faint photometric sequences in compact fields suitable for calibrating
modern equipment.  They presented {\it BVRI\/} magnitudes for some 800 stars in
the fields of three globular clusters and, with one exception, found generally
good agreement with previously published photometry for the same stars.  The
exception was the $I$-band magnitudes in the cluster \ngc{2419}, where they
found a systematic offset compared to my posted data.  The principal purpose of
the present paper is to investigate the origins of this discrepancy.

\section{\ngc{2419}}

The distant globular cluster \ngc{2419} is a fascinating target in its own
right.  According to the summary of cluster properties compiled by
William~E.~Harris\footnote{
http://physwww.physics.mcmaster.ca/\%7Eharris/mwgc.dat} (Revision: February
2003), the cluster lies at the position $\alpha\,=\,07$\hr38\mn08\Sc5,
$\delta\,=\,+35$\deg52\min55\sec (J2000), $l^{II}\,=\,$180\deg,
$b^{II}=+25$\deg, thus in the Galactic anticenter direction and somewhat above
the plane.  It is some 84$\,$kpc from the Sun and 92$\,$kpc from the Galactic
Center, making it the fifth most remote globular cluster believed bound to the
Galaxy, surpassed only by Eridanus (95$\,$kpc from the Center), Palomar~3
(96$\,$kpc), Palomar~4 (109$\,$kpc) and AM$\,$1 (122$\,$kpc).  Indeed,
among the eleven dwarf galaxy companions of the Milky Way, seven lie closer to
the Galaxy than NGC$\,$2419 (Sagittarius, LMC, SMC, Ursa Minor, Draco, Sculptor,
Sextans), and only four lie beyond it (Fornax, Carina, Leo~I, Leo~II).  
The recently discovered system SDSSJ1049+5103 (Willman \etal\ 2004) lies at
roughly half the distance of \ngc{2419}.

Unique among the outer-halo globulars, \ngc{2419} is extremely luminous---with
$M_V \approx -9.6$---and is tied with \ngc{6441} for third place among the
globular clusters of the Milky Way, after \ngc{6715} (--10.0) and
$\omega$~Centauri = \ngc{5139} (--10.3).  None of the other globular clusters in
beyond-the-SMC space (the aforementioned four plus Palomar~14) is brighter than
$M_V\approx-6$ (Pal~4), or roughly 3\% of the luminosity of \ngc{2419}. 
However, like the other outer-halo globular clusters, \ngc{2419} has a core
radius much larger than is found among the closer-in globulars:  its half-light
radius is some 18$\,$pc, while most globular clusters closer than the SMC
have half-light radii in the range 2--9$\,$kpc.  This combination of high
luminosity, large radius, and great distance have led van~den~Bergh \& Mackey
(2004) to propose that \ngc{2419} is in fact the stripped core of a former
nucleated dwarf spheroidal galaxy (an origin also proposed for
$\omega$~Centauri, which also has an extremely high luminosity and an unusually
large radius, see Bekki \& Freeman 2003 and references therein).  

Harris \etal\ (1997) analysed images of \ngc{2419} obtained with WFPC2 on the
{\it Hubble Space Telescope\/} as part of a program to investigate the age
structure of the globular cluster population.  They found that in the morphology
of its color-magnitude diagram, \ngc{2419} closely resembles the inner-halo
cluster M92 = \ngc{6341}, which it also closely resembles in metal abundance
(\ie, [Fe/H] = --2.12 for \ngc{2419}, --2.28 for M92 according to the Harris
compilation catalog; see Harris \etal\ for a more detailed discussion).  The
ages of the two clusters were found to be identical to within the precision of
the photometry.

In large part because it was included among the objects targeted by the ``KPNO
Video Camera/CCD Standards Consortium'' (Christian \etal\ 1985), \ngc{2419} has
more data in the various imaging archives around the world than your average
pointing on the sky.  Many assiduous astronomers who have no interest at all in
the cluster {\it qua\/} cluster have nevertheless observed it in hopes of better
calibrating their data for objects they truly care about.  Most of those
astronomers included other Landolt and Consortium standard fields on their
observing programs for the same reason.  Of course, a few other astronomers
have observed \ngc{2419} because they wanted to study {\it it\/} specifically
and, in general, they have also observed other standard fields on the same
nights.  

Because the target is so far away and because it has a very low metallicity, the
giant branch and the blue horizontal branch provide an extremely broad color
baseline at a magnitude level which is both bright enough for 2m-class
telescopes and faint enough for {\it HST\/} and 8m-class telescopes with
sensitive CCDs.  These facts make \ngc{2419} among the most tempting targets in
the sky for the establishment of a faint photometric sequence.  This, plus the
fact that \ngc{2419} had already been observed with WFPC2 on {\it HST\/} for the
Harris \etal\ study (GO Proposal \#5481: ``Ages for the Outermost Globular
Clusters: The Formation of the Galactic Halo: Cycle 4 High,'' PI Hesser) led
directly to its inclusion in the target lists for HST proposals 6937: ``WFPC2
Cycle 6 CTE Calibration,'' Stiavelli; 7628: ``WFPC2 Cycle 7 Photometric
Characterization,'' Whitmore; 7630: ``WFPC2 Cycle 7 CTE Characterization,''
Casertano; 8821: ``Cycle 9 WFPC2 CTE Monitor,'' Riess; 9043: ``Cepheid Distances
to Early-type Galaxies,'' Tonry; 9591: ``Cycle 11 WFPC2 CTE Characterization,''
Whitmore; and 9601: ``WFPC2-ACS Photometric Cross-Calibration,'' Koekemoer. 
Observations of \ngc{2419} were also obtained with ACS in the course of program
9666: ``Photometric Transformations,'' Gilliland.  In each case, \ngc{2419} was
observed specifically for the value of its local standard-star sequence in
calibrating the photometric performance of the {\it HST\/} cameras.  The legacy
value of the Hesser/Harris data was also exploited by program 8095: ``Accurate
proper motions of Galactic halo populations,'' Ibata.

It is obvious that certain, accurate knowledge of the Landolt-system magnitudes
and colors of stars in the field of \ngc{2419} would be of extreme value not
only for understanding the idiosyncracies of that one astronomical object, and
not just for somewhat better calibration of ground-based photometric images,
but also for helping determine just how much confidence one can place on
photometry obtained with a multi-billion-dollar space observatory.

\section{REDUCTION METHODOLOGY}

The use of DAOPHOT, ALLSTAR, ALLFRAME, DAOGROW, etc.\ to produce instrumental
magnitudes from digital CCD images has been thoroughly discussed elsewhere.
Here I will concentrate on how the aperture-corrected instrumental magnitudes
are referred to the standard photometric system.

To describe the provenance of the present photometry, I must explain a few
terms.  The meaning of the word {\it photometric\/} as regards a night of
optical observations is familiar to most astronomers.  In brief, it is a night
upon which the extinction of incoming starlight is a nearly linear function of
the path-length through the terrestrial atmosphere (when flux is expressed in
logarithmic units, such as magnitudes) and is at most a weak function of time or
azimuth.  A {\it non-photometric\/} night is one where useful amounts of
starlight are able to reach the telescope, but the assumptions of temporal and
directional uniformity break down.  

In mathematical terms, on a photometric night I relate the observed instrumental
magnitudes to the standard magnitude system through an equation of the form
$$O_{ij} = L_i + Z + A\cdot C_i + K\cdot Q_j + \ldots \eqno(1)$$
where $O_{ij}$ is the observed instrumental magnitude of standard star $i$ in
CCD image $j$ ($O \equiv 25 - 2.5\log[\hbox{\rm DN sec}^{-1}]$); $L_i$ is the
standard magnitude of the same star on the system of---in the present
instance---Landolt (1992); $C_i$ is one of the usual colors of the star in the
same standard system (\eg, \bmv, \vmi, or \umb\ as appropriate); $Q_j \equiv
X_j-1$, where $X_j$ is the airmass of the $j$-th CCD image; and $Z$, $A$, and
$K$ are taken to be constants for an entire {\it dataset\/} (to be defined
below).  In this formulation, $Z$ represents the standard-to-observed correction
for a star of zero color observed at the zenith.  

The ``\ldots'' in Eq.~(1) represents the fact that the software permits
additional arbitrary polynomial terms in the various standard-system colors $C$,
in $Q$, in $R \equiv Q\cos({\rm azimuth})$, in $S \equiv Q\sin({\rm azimuth})$,
in $T \equiv$ the time of the observation in ($\pm$) hours from midnight UT, and
in the $x$ and $y$ coordinates of the stars measured in the natural system of
the CCD.  The terms in $R$ and $S$ provide the ability to allow for modest
directional nonuniformity of the extinction (such as the different
meteorological effects of an ocean to the west and high desert mountains to the
east, for instance), and are both continuous and smooth passing through the
zenith---which is why I decided to use $Q\equiv X-1$ rather than $X$ to
parameterize the airmass.  Similarly, polynomial terms involving $T$ provide
the flexibility to model small temporal variations in the extinction.  Finally,
the terms in $x$ and $y$ permit the correction of variations in photometric
zero-point across the face of the camera due to scattered light in the flat
fields or other illumination problems in an imperfect camera system. 
Calibration terms may be formed from any product of powers of colors, $Q$, $R$,
$S$, $T$, $x$, and $y$ as may be required by the data.  For instance, for the
$B$ photometric bandpass I usually include a term in $(\bmv)\cdot Q$ with an
empirically determined coefficient of --0.016 to allow for the fact that the
light of blue stars experiences more extinction than the light of red stars when
passing through the terrestrial atmosphere, since more of their flux is
concentrated toward the short-wavelength side of the bandpass.  (Similar terms
in the {\it VRI\/} transformations have been tried, and been found to be
statistically insignificant.)  I also generally include terms in $C^2$ for the
broad {\it BVRI} photometric bandpasses.  

The transformation model on a non-photometric night is slightly different:
$$O_{ij} = L_i + Z_j + A\cdot C_i + \ldots \eqno(2)$$
In this case, each CCD image is assigned its own photometric zero-point, $Z_j$,
on the basis of the standard stars it contains, while the color transformation
is considered a constant to be determined for the dataset as a whole.  Any CCD
image containing two or more standard stars with a significant spread of
standard-system color contributes to the definition of the color transformation
as well as its own zero-point.  A CCD image containing a single photometric
standard can be assigned a photometric zero-point, but it contributes nothing to
the definition of the color transformation; on a non-photometric night, an image
containing no standards is useless for any photometric purpose (although such an
image will be included in the ALLFRAME reductions for the contributions it can
make to the completeness and astrometric precision of the star list).  On a
non-photometric night, transformation terms in $Q$, $R$, $S$, and $T$ are
generally superfluous, although higher-order terms in color and terms in $x$ and
$y$ can still be determined when there are enough standard stars in at least
some images to allow their unambiguous determination.

A {\it dataset\/} is that block of data which is included in a single solution
of a set of transformation equations---one for each filter---of either form (1)
or form (2).  A dataset is usually either the data from a single photometric
night or the aggregate of the data from several non-photometric nights from the
same {\it observing run\/}.  An observing run is a set of consecutive or
near-consecutive nights with the same telescope/camera/filter/detector system
with no (to the best of my ability to judge, in the case of archival data)
intervening instrument changes.  

When several datasets are available from the same observing run (\ie, several
distinct photometric nights, or a mix of photometric and non-photometric
nights), I begin by performing completely independent photometric reductions for
all datasets.  I then intercompare the resulting values for the transformation
coefficients from the different samples.  If they are inconsistent, I attempt to
identify the cause, which is usually too few standard-star observations with a
too-small range of airmass and/or color in one or more of the datasets.  

The reductions of the different datasets from a given observing run may then be
partially coupled together, as follows.  The weighted average value of some
coefficient in the transformation for each filter, generally beginning with the
highest-order color term, is calculated from those datasets where it is well
determined.  It is then imposed as a ``known'' constant on all the datasets from
the run, which are then re-reduced with one unknown fewer per bandpass.  This
process is repeated until only those coefficients which appear to be
legitimately different from one dataset to the next---such as the linear
extinction coefficient and the zero-point---remain freely and independently
determined for each dataset.  (Since the zero-point is defined at the zenith
rather than outside the atmosphere, real changes in the extinction are reflected
in real changes in the zero-point.  This is not a problem for my approach.)
Occasionally the night-to-night differences are sufficiently small compared to
the observation-to-observation scatter that it makes sense to impose unique
zero-points and extinction coefficients on several consecutive photometric
nights, making them in effect a single dataset.

The data from different chips of a multi-CCD mosaic camera are always treated
as coming from distinct cameras and, hence, as constituting different datasets. 
That is to say, before obtaining the instrumental photometric indices no attempt
is made to patch or ``drizzle'' the individual images into a larger master image
of the field, and PSFs and aperture corrections are defined independently with
no attempt to force continuity between adjacent chips.  If it is judged that a
variable PSF is required (in practise, I always employ variable PSFs when
reducing data from large-format cameras), the variation is expressed in terms of
the natural $(x,y)$ coordinate system of each chip without reference to the
position of the chip with respect to the optical axis of the telescope/camera
system.

At the stage of calibrating the instrumental photometry to the standard system,
however, the results of a single exposure from the different CCDs of a mosaic
may be partially coupled, just as the separate nights from a single observing
run may be coupled.  In particular, after an initial reduction of the data
from a photometric night, average extinction coefficients will generally be
calculated and imposed in common to the different chips of the mosaic for a
subsequent reduction run.  Furthermore, if the data do not clearly demonstrate
otherwise, often the high-order and occasionally the linear color terms in the
transformations will be imposed as constant for all chips.  The spectral
response of a telescope and camera is determined by the transparency of the
atmosphere; the reflectivity of the mirror(s); the transmissivity of the
corrector, the dewar window and any other refractive optical elements; the
throughput of the filter; and the spectral response of the CCDs.  All of these
components except the last are common to all the chips, and if the CCDs are from
the same batch (and if the filter is spatially uniform), real chip-to-chip
differences in the shape of the spectral response may not be perceptible at the
level required here.  The different chips in a mosaic camera are always
permitted to have individual photometric zero-points.  

Occasionally, when observations for only a handful of standard stars are
available and the range of airmass spanned is small, I will couple the
extinction coefficients in the different filters.  Average values formed
from many excellent photometric nights with hundreds of standard stars
on Kitt Peak, Cerro Tololo, La Silla, and La Palma consistently give extinction
ratios $K_U : K_B : K_V : K_R : K_I$ close to $1.91 : 1.00 : 0.56 : 0.42 :
0.35$.  For instance,  typical extinction coefficients for 7,000-foot
observatories are $K_U = 0.48$, $K_B = 0.25$, $K_V = 0.14$, $K_R = 0.105$, and
$K_I = 0.09\,$mag~airmass${}^{-1}$.  On nights when such cross-filter averaging
seems appropriate, the measured extinction coefficients in whichever filters are
available are scaled to the $B$-equivalent extinction using these same ratios, a
weighted average is formed, and this is scaled back to the appropriate values
for the individual filters.  I am convinced that this method is less likely to
result in significant systematic errors than permitting the individual
extinction coefficients to take on values entirely {\it ad libitum\/},
especially when standards are few and the range of airmass is small.  On a few
nights where the range of airmass is effectively nil, I impose mean extinction
coefficients for the site rather than attempting to determine them from the
data.  

Once the effective mean zero-points for a night of observations have been
established from the standard stars, inaccurate color transformations or
extinction laws, and failure to recognize and correct any residual dependences
on time of night or position on the detector will impose positive errors on the
derived magnitude estimates from some observations and negative errors from
others.  Therefore, when averaged over many different runs, telescopes, and
detectors, such neglect should primarily amplify the random noise in the
averaged results rather than introducing major systematic errors in the final
photometric catalog.  Suspicion can and should still adhere to ``standard''
stars established in only one or only a few observing runs.

\section{DATA}

The current generation of photometric indices for my standard stars is based
upon an analysis of 649 datasets resulting from 622 nights of observations
spread over 179 observing runs.  Of these, 516 datasets from 336 nights were
considered photometric and the remaining 133 datasets from 286 nights were
reduced in non-photometric mode.  (The number of photometric datasets exceeds
the number of photometric nights because they include data from mosaic 
cameras; the number of non-photometric datasets is less than the number of
non-photometric nights because I often merged the data from consecutive
non-photometric nights into single datasets.)  

Table~1 lists those observing runs where data for \ngc{2419} were obtained.  The
observing-run labels in the first column are arbitrary and occasionally
whimsical, so great significance should not be attached to the nomenclature. 
The second and third columns identify the telescope and detector that were used
to acquire the data, and the fourth column indicates the approximate dates of
the observations.  The columns labeled ``Clr'' and ``Cld'' indicate the number
of {\it datasets\/} resulting from the observing run that were reduced in
respectively, clear-weather (Eq.~1) and cloudy-weather (Eq.~2) mode.  Each of
the MiniMos runs consisted of a single night of observing, but since the camera
contains two CCDs, each night produced two datasets.  Finally, the last four
columns list the number of individual {\it exposures\/} of \ngc{2419} that were
obtained in each of the {\it BVRI\/} filters.  In the case of the MiniMos
observations, the actual number of CCD images is twice the number of exposures;
since the chips have no overlap on the sky, the maximum number of images that a
given star can appear in is given by the number of exposures rather than the
number of images.  The observing runs listed for the WIYN 3.5m telescope and the
WIYN 0.9m telescope run labeled ``abi36'' are the ones that were analysed by
Saha \etal\ (2005), with the exception of the run labeled ``wiyna.''  This night
was non-photometric and was not included in the Saha paper, but I include it
here, having analysed the night's data using the non-photometric reduction mode
described above.  I include the night labeled ``wiynb'' here for completeness,
even though no observations of \ngc{2419} were made on that night, because it
will contribute to many of the comparisons that follow.

It should be noted that the images from the WIYN telescopes---both the 3.5m and
the 0.9m---were bias-subtracted, flat-fielded, and otherwise preprocessed by
Abi~Saha before he gave them to me.  Any differences of opinion on the
quantitative results of these observations will originate in the analysis, not
in how the data were acquired or preprocessed.

\section{ANALYSIS}

The derivation of the transformation constants from standard-star observations
and the application of those transformations to program stars is carried out by
two separate programs (Stetson 1993).  CCDSTD considers all stellar observations
in a given dataset, and solves Eqs.~(1) or (2) using a robust least-squares
analysis to determine the desired quantities $Z$, $A$, and $K$, etc., by
comparing the observed quantities $O$ to the ``known'' quantities $L$, $C$, and
$Q$.  If the universal time, the azimuth of the observation, or the $(x,y)$
position of the star on the chip are included in the transformation, they are
also considered to be error-free, known quantities.  The instrumental magnitudes
on the left-hand sides of the equations are considered to be measured quantities
with associated standard errors; the errors assigned include both the measuring
errors for the instrumental magnitudes as derived from the analysis of the CCD
image and the standard errors of the mean magnitudes given along with the
photometry in the published standard-star lists.  This approach is
mathematically valid as long as the statistical uncertainty in the $O_{ij} -
L_i$ differences overwhelm the uncertainties in $A\cdot C$, $K\cdot Q$, and the
azimuth, time, and position effects.

The transformations thus derived are applied to the program stars by a different
computer program, which employs the same two equations in a reversed sense. 
CCDAVE considers the data for only one star at a time, but it analyses
simultaneously all the observations of that star from all datasets.  It assumes
that the values $Z$, $A$, $K$, and the other transformation parameters
appropriate to each dataset are now known quantities.  The instrumental
magnitudes $O$ on the left side of the equations are still measured quantities
with associated uncertainties, and it is the standard-system magnitudes, $L$,
which are now the unknown parameters to be determined by robust least squares. 
This requires some iteration, since the stars' colors are not initially known. 
As each star is reduced its color is first assumed to be zero.  Then all of the
Eqs.~(1) and (2) for all observations of that star from all filters, all nights,
all observing runs, all telescopes, are solved by robust least squares to derive
the unique maximum-likelihood standard-system magnitudes.  These are used to
compute the appropriate colors, and the least-squares reduction of the system of
equations is repeated until the derived magnitudes and colors stop changing, a
process which normally requires only a few iterations.  These magnitudes and
their associated standard errors are written to a computer file and the software
proceeds to the next star on the list.

In what follows, I will attempt to adhere to the following naming scheme.  If I
refer to the ``Landolt (1992)'' photometric system, I am speaking of the
photometric indices and standard errors given in that specific
publication---although see the Appendix for some modest adjustments I have made
to Landolt's published standard errors.  The name ``Landolt system,''
unmodified, will mean that in the {\it UBV\/} bandpasses I have formed the
weighted average of results taken from Landolt (1992) and Landolt (1973)
after the latter have had small ($\ltsim\,0.002\,$mag) zero-point adjustments
applied; these offsets have been determined by direct comparison of the
results for  stars common to the two publications.  The ``Landolt system'' is
also augmented by a very few stars from Landolt (1983) which did not reappear in
the 1992 publication.  ``Present photometry'' or ``my photometry'' will refer to
the net results of my current analysis of the 649 datasets considered here,
individually referred to the Landolt system as accurately as I can do it.  These
data, averaged together with Landolt's for stars in common,  were posted to my
Web site on January 27, 2005.  However, in what follows ``the posted
photometry'' or some equivalent phrase will refer to my photometric
indices derived from 594 datasets as of September 28, 2004 (likewise averaged
with Landolt's for stars in common).  These are what were available through my
Web site when Saha \etal\ submitted their paper.  ``Primary standards'' refers
to the average of my results with Landolt's for stars in his standard lists;
``secondary standards'' refers to stars not observed by Landolt that satisfy my
acceptance criteria based on my data alone.

\subsection{Was the posted photometry on the Landolt system?}

The short answer is, not entirely, although I thought it was.

CCDAVE---the program that applies the derived calibration equations to convert
instrumental-system magnitudes to the standard system---is not given any prior
information about the star's expected standard-system magnitudes.  It is
provided with only the measured instrumental CCD magnitudes and the numerical
values of the coefficients in the various datasets' transformation equations. 
Therefore I can test the software by giving it the instrumental magnitudes for
the standard stars as if they were program stars.  If CCDAVE returns values very
close to to the tabulated standard values for those stars, then it means that
CCDSTD has adequately evaluated the coefficients in the transformation equations
and has thus defined a good mapping between the instrumental and standard
systems.  It also means that CCDAVE is capable of correctly applying those
transformations in the reverse sense.  A failure of {\it either\/} either of
those steps would make it extremely unlikely for the output of CCDAVE to agree
with the input to CCDSTD for most stars.  Since CCDAVE has no way of knowing
whether a given star is a standard or a program object, if the transformations
and the software work correctly for the standards they should also work
correctly for the rest.  

In the past, I have used plots like that presented here as Fig.~1 to
satisfy myself that my software and derived coefficients were adequate to 
transform the available instrumental magnitudes for Landolt's stars to his
standard system (\cf, for instance, HP-II, Figs.~1--4; HP-III, Figs.~2 and 3). 
The present figure compares the results of my analysis (as of September 28,
2004) of the observational data for Landolt's standards from 594 datasets
to Landolt's published magnitudes for the same stars.  In the figure I have
plotted the {\it posted\/} minus {\it Landolt\/} magnitude differences in $B$,
$V$, $R$, and $I$ versus \bmi\/ color for all stars meeting the following
conditions: at least four Landolt observations and a standard error of the mean
Landolt magnitude not greater than 0.03$\,$mag, {\it and\/} at least four
observations under photometric conditions and a standard error of the mean
magnitude not greater than 0.03$\,$mag in my results, {\it and\/} no evidence
within my data of intrinsic variability in excess of 0.05$\,$mag, r.m.s. 
Photometric differences are plotted for 230 stars in $B$, 261 in $V$, 144 in
$R$, and 163 in $I$, and the error bars represent a $\pm\,1\sigma$ standard
error of the mean centered on the observed difference.  The \bmi\ colors are the
average of Landolt's and my results, and the error bars include both Landolt's
tabulated standard errors and my own added in quadrature.  The standard
deviation of the magnitude residuals is 0.015$\,$mag in $B$, 0.012$\,$mag in
$V$, 0.017$\,$mag in $R$, and 0.017$\,$mag in $I$, which are all larger than
would be expected from the claimed measurement uncertainties.  In order to
explain the amount of scatter in these magnitude differences, I have to assume
that a ``cosmic'' star-to-star scatter is present in addition to the random
measurement uncertainties claimed by Landolt and me: this cosmic scatter is
0.010$\,$mag in $B$, 0.008$\,$mag in $V$, 0.011$\,$mag in $R$, and 0.013$\,$mag
in $I$.  These additional dispersions are not included in the error bars shown
in the figure.  

These data suggest that my magnitudes are on the Landolt system to a high
level of accuracy:  the absolute value of the weighted mean difference is $\leq
0.6\,$mmag and the standard error of the mean difference is $\leq\,1.4\,$mmag
in all four filters.  In preparing the photometry for posting on the internet,
my next step has always been to apply these mean magnitude differences ($\leq
0.6\,$mmag in the present instance) as additive offsets to my derived
standard-system magnitudes to place them truly on the Landolt system in the
mean.  I then take weighted averages of Landolt's and my magnitudes for each
star in common, and these are made available to the community.  However, there
is no magic about requiring at least four observations and no more than
0.03$\,$mag of uncertainty;  I could as easily compare, for instance, stars
with a minimum of eight measurements and no more than 0.02$\,$mag of uncertainty
in each sample.  In this case, I find a mean magnitude difference in $V$ of
+0.6$\,$mmag based on 174 stars, and a difference in $I$ of --1.0$\,$mmag based
on 116 stars.  Conversely, if I accept stars with two observations and
uncertainties of 0.05$\,$mag, the differences are --0.3$\,$mmag (383 stars) and
--3.0$\,$mmag (235 stars).  Similar changes to the comparison could be made by
choosing a different but still reasonable weighting scheme: for instance, by
assuming a common cosmic dispersion of 0.010$\,$mag in all four filters.  Since
the selection criteria for acceptible standards and the details of the weighting
scheme are arbitrary, it is not clear that ``Landolt's system'' is even defined
to a level better than a few millimagnitudes, and it is certainly not 
meaningful to claim to be on it to an accuracy better than that.  

Unfortunately, this turns out not to be the whole story.

Spurred by the Saha \etal\ paper, I investigated more deeply.  In particular, I
binned these magnitude differences for stars in intervals of 0.50$\,$mag of
\bmi\ color, and determined the unweighted median difference within each bin.
The results are shown in Fig.~2.  Here the error bars represent a robust measure
of the dispersion of the magnitude differences within each bin:  $\sqrt{\pi/2}$
times the mean absolute deviation.  If the distribution of the magnitude
differences were Gaussian in form, this would equal the standard Gaussian
$\sigma$.  Please note that this represents the spread of the residuals in each
bin; in most cases error bars representing the standard error of the median
difference in a bin would be smaller than the dot itself.  (Note also that the
reddest bin contains only a single star---Landolt~110~273---and I have no
acceptable $B$ magnitude for this star.  Landolt observed this star only five
times, so its data are not much better than the minimum acceptable quality: if
my acceptance criteria were only a little more stringent, this star would not be
included.  My own results for it are highly concordant:  the r.m.s.\ 
dispersion in the measured magnitudes, above and beyond the estimated random
measuring errors, is only 0.004$\,$mag based on a total of 170 individual
measurements in the various filters.)  It is evident that there is a substantial
trend in the $I$-band residual as a function of a star's color, with perhaps
weaker trends in the other photometric bandpasses.  

Evidently color-scale errors have crept into my broad-band magnitude systems
as, over the years, I have transformed various observing runs' data to the
Landolt system, averaged my results with Landolt's for stars in common and
created new secondary standards based on my data alone, used the expanded
standard list to recalibrate existing datasets, and added new datasets to the
expanding corpus of observations.  I have no idea when or how the error
occurred; one possibility is a typographical error in an earlier version of my
hand-typed catalog of Landolt's standard indices; I have discovered and
corrected a few such blunders over the years.  (The use of robust statistical
methods prevents such gross errors from having a disastrous effect on the final
results---a disastrous effect would have revealed the mistake---but much more
subtle effects, such as this one, {\it might\/} still have resulted from such an
error.) However it entered, since the total number of my photometric
measurements now greatly exceeds the number of Landolt observations, such a
scale error---once established in my magnitude system---became self-sustaining. 
In the case of $\Delta I$ versus \bmi, a more sophisticated fitting procedure
applied to the data indicates that the size of the effect is 0.007$\,$mag
mag${}^{-1}$, as illustrated by the dashed line in the bottom panel of
Fig.~2.  A similar, perhaps weaker, trend also appears to be present in $V$ and
possibly in the other filters.

For the rest of this paper and for the current and future releases of my
standard-star data, I will remove any net offsets and trends with color from my
results by brute force as follows.  Using only those stars meeting the
above-stated acceptance criteria (four observations and standard errors of the
mean magnitude $\leq\,0.03\,$mag in {\it both\/} samples, and no evidence of
variability in excess of 0.05$\,$mag) I average Landolt's and my results and
calculate all possible simple color indices involving a given bandpass.  For
instance, I carry out a weighted fit of the differences $\Delta I$ to low-order
polynomials in as many of \umi, \bmi, \vmi, and \rmi\ as are defined, given the
above selection constraints.  (In all my work to date, I have carried $U$-band
magnitudes along in the process whenever available.  However, I have not
released these results to the community because I do not believe them to be
competitive with Landolt's published indices.)  Then for each star, an
unweighted average of the available predicted values of $\Delta I$ is applied to
my derived result to place it more accurately on the Landolt system.  I must
consider all possible colors because not all bandpasses are available for all
the secondary standard stars; for any given star I compute the mean correction
based upon the indices that I have.  Similar corrections are derived and applied
to the other photometric bandpasses.

After torquing my results back to the Landolt system, I freed any color
terms that had previously been fixed for particular datasets, and re-ran CCDSTD
to compute new transformations for all the observed data.  Fig.~3 shows
the current bin-averaged residuals that result from the latest round of
reductions for 649 datasets.  Whatever the remaining residual differences
between my magnitude systems and Landolt's it is clear that they are not easily
described by low-order polynomials in color.

Between September 2004 and January 2005, I have added 55 additional datasets to
my sample---including two with observations of \ngc{2419}---and defined 2,124
additional standard stars in various fields.  These have been added to the
body of data used to retroactively recalibrate my standard sequences, including
that in \ngc{2419}.  Fig.~4 plots the resulting differences between the
January 2005 and September 2004 values of the $V$ and $I$ magnitudes of the
\ngc{2419} sequence on a star-by-star basis, in the sense (January 2005) minus
(September 2004), versus \vmi\ color.  Eeach error bar represents a $\pm 1$
standard error of the mean magnitude of a particular star at {\it either\/}
epoch:  that is, the September 2004 estimated standard errors and the January
2005 standard errors have been added in quadrature and divided by $\sqrt2$
to obtain a representative uncertainty for each star.  Here a systematic linear
change in the $V$-magnitude system is also evident, with an amplitude similar to
that in $I$.  This means that my \vmi\ colors for stars redder than $\vmi
\approx 0.5$ have hardly changed.  The correction applied here to eany indivdual
star consistently exceeds my claimed standard errors only for stars redder than
\vmi$\,\sim\,$1.5, although I admit that this is comparing a systematic effect
to random uncertainties:  the magnitude {\it system\/} has certainly changed by
more than my previous estimate of its uncertainty.  

Considering the standard system as a whole, including stars bluer than the
bluest of the \ngc{2419}\ sequence, the systematic change reaches extreme values
of $\Delta I = -0.017\,$mag and $+0.022\,$mag for stars of extreme color,
$\vmi\sim -0.35$ and $\vmi \sim 2.65$, respectively (corresponding roughly to
\bmv\ values of --0.3 and +2.3, or stars bluer than spectral class B0 and redder
than M8).  The absolute value of the correction is $< 0.005\,$mag for the most
common stars with $0.6 < \vmi < 1.4$ ($0.5 < \bmv < 1.2$), and within the same
limits the r.m.s.\ correction is about 0.003$\,$mag.  This latter color range
contains more than half (57\%) of the Landolt standards and presumably a larger
fraction of nearby stars, since Landolt made a particular effort to include
stars with extreme colors in his sample.  Furthermore, when stars of all colors
are considered, the root-mean-square systematic error due to the erroneous color
slopes in the September 2004 posted data amounts to a bit less than
0.008$\,$mag; this is smaller than the $\sim\,0.013\,$mag cosmic star-to-star
scatter in the $I$-band differences found by Stetson, Bruntt \& Grundahl and
found again above in this paper.  Therefore this is not the main component
of the scatter that we attributed to ``bandpass mismatch,'' which continues to
dominate the Stetson {\it versus\/} Landolt differences over nearly all of the
color range.  It should also be remembered that my posted magnitudes for the
{\it primary\/} standards were the average of my results with Landolt's, so for
researchers employing those stars, at least, the effect of my calibration error
would have been further diluted.  My secondary standards, however, included the
mistake at full strength.  

The color-slope errors in my September 2004 posted data cannot be the sole, or
even the main, explanation for the difference between my $I$-band magnitudes and
those of Saha \etal, which were $\sim\,0.04\,$mag in the middle color range
where the present correction has negligible effect.  Abi Saha has very kindly
provided me with electronic copies of his main data tables.  The $(x,y)$
coordinates included in them enable me to match my stars up with his on a
star-by-star basis.  Fig.~5 shows the $V$-band and $I$-band magnitude
differences for the NGC2419 stars in the sense (posted September 2004) minus
(Saha) as a function of \vmi\ color.  The $\pm1\sigma$ error bars are the result
of adding Saha's estimated standard errors in quadrature with mine, which
somewhat overestimates their size: since the data I have analysed for \ngc{2419}
includes all the images used by Saha \etal, any error sources that are intrinsic
to the images themselves (\eg, photon noise and flat-fielding errors) will have
been included twice.  However, I have no way of knowing what fraction of the
error budget comes from those sorts of problems, and how much from differences
in the analysis, such as the estimation of the sky brightness, the removal of
the effects of neighboring stars, and the treatment of the transformation from
instrumental to standard magnitudes.  Therefore, all I can say is that the error
bars should be somewhere between 0.7 and 1.0 times the length shown; a
factor near 1.0 is much more probable than one near 0.7.

In comparison, Fig.~6 shows the \ngc{2419}\ magnitude differences for (January
2005) minus (Saha).  In both plots, I have included all stars with at least four
measurements and $\sigma(\hbox{magnitude}) \leq 0.03\,$mag in both datasets.  In
each filter the main ridgeline of the data has rotated so that the differences
now pass through zero near $\vmi\sim2.0$ and the center of gravity of the bluest
stars is now slightly less positive, compared to before.  However, the net
offset of $\Delta I \sim\,0.02$--$0.04\,$mag for the vast majority of stars in
the center of the color range has not changed much.  Taking the sample as a
whole, my calculated mean differences with respect to Saha \etal\ are
--0.0092$\,$mag in $V$ and --0.0268$\,$mag in $I$ for the September 2004
results, and --0.0049$\,$mag in $V$ and --0.0211$\,$mag in $I$ for the January
2005 results.  My previous color-scale error thus accounts for about
0.005$\,$mag in each filter, or half the mean discrepancy in $V$ and 20\% of the
mean discrepancy in $I$ averaged over all colors.  

\subsection{Is the present photometry for \ngc{2419} on the standard system?}

To transform the instrumental magnitudes measured for the program stars to the
standard system, I employ precisely the same lines of code that successfully
transform instrumental magnitudes to standard-system magnitudes for the stars in
Landolt's list.  If there were any systematic errors due to intrinsic
differences between the secondary standards and the Landolt standards (for
instance, an unrecognized photometric nonlinearity in the detector operating on
the comparative faintness of the program stars relative to the Landolt
standards) these should tend to average out over the 649 different observational
datasets.  The systematic error would then show up when the data from an
offending run are compared to the average of all runs.  I continually look for
such effects as I reduce the data.  To survive as a serious problem, any such
systematic error would have to be consistent across many observing runs using
many cameras on many telescopes.

To investigate this issue I have compared my present results from the 616
datasets that do not contain any observations of \ngc{2419} on the one hand to
the 33 datasets that do contain observations of \ngc{2419} on the other.  Note
that any given observing run could potentially contribute datasets to both of
these subsamples if \ngc{2419} was observed on some nights of the run and not
on others.  Again I consider all stars (Landolt standards and secondary
standards) with at least four observations under photometric conditions and
standard errors of the mean magnitude not greater than 0.03$\,$mag in {\it
each\/} of the two subsamples, as well as no evidence of intrinsic variability
in excess of 0.05$\,$mag, r.m.s., when all observations are considered
together.  A total of 1,897; 3,587; 1,807; and 3,343 stars meet these selection
criteria in $B$, $V$, $R$, and $I$.  Note that the \ngc{2419} standard sequence
itself is {\it not\/} included in this comparison since by design the first
sample contains no observations of those stars.  The comparison is based on the
{\it other\/} primary and secondary standard stars that have been observed on
the same nights and in the same manner as the \ngc{2419}\ secondary standards.

The standard deviation of the differences in the mean magnitudes of the
without-2419 versus with-2419 datasets is 0.017$\,$mag per star in $V$ and
0.020$\,$mag in $I$.  To explain the size of these differences I have to assume
that a star-to-star cosmic scatter of order 0.010$\,$mag in $V$ and 0.012$\,$mag
in $I$ is present in addition to the estimated uncertainties of the raw
magnitude measurements.  As discussed before, these could be the result of
subtle differences in spectral energy distributions as perceived by different
filter/detector combinations.  The mean differences between the not-2419 and the
2419 datasets are $+1.3 \pm 0.3\,$mmag in $V$ and $+3.8 \pm 0.3\,$mmag
in $I$, in the sense that the results of the with-2419 runs are brighter in both
$V$ and $I$.  I was initially surprised by the size of these differences,
but I suppose they are still acceptable.  {\it My experience is that the results
for any given field from any given night of CCD observations can be
systematically in error by $\sim\,0.02\,$mag.}  Only 24 of the 33 \ngc{2419}
datasets were reduced in ``photometric weather'' mode (Eq.~1) and so were
capable of constraining the absolute zero-points of the magnitude scales, and of
these only 16 included $I$-band data: $0.02/\sqrt{16} \approx 5\,$mmag, so the
not-2419 versus 2419 difference of 3.8$\,$mmag in $I$ is still within the
expected standard error of the mean.  

The previous section has shown that the inherent scatter in star-by-star plots
can mask real trends in the data.  Therefore, Fig.~7 plots the
{\it median\/} magnitude differences between the 616 datasets that do not
contain observations of \ngc{2419} and the 33 datasets that do.  The stars have
been divided into 1.0-mag intervals of $V$ magnitude and the unweighted median
magnitude differences in $B$, $V$, $R$, and $I$ have been plotted against the
mean $V$ magnitude in each bin.  Appreciable differences are seen in the faintest
two bins in $B$, which contain, respectively, eight and three stars that must
have been near the detection limit in many of the datasets where they appear. 
The brightest points in $B$ and $I$ are also off, but this bin contains
precisely one star, Landolt's 98~653 at $V = 9.53$, which may have been close to
saturation in many of the frames included in one sample or the other.  Apart 
from those exceptions the plots are quite flat, rarely deviating from zero
by as much as 0.01$\,$mag over a range of 10 or 11 magnitudes.

The data for the $I$ bandpass do suggest that there may be a slight difference
in the magnitude systems for the \ngc{2419}\ nights as compared to the non-2419
nights.  Only 355 of the 616 non-2419 datasets included observations in the $I$
filter and, as already noted, only 26 of the 33 datasets including observations
of \ngc{2419} had $I$-band data.  Assuming that the average results from 355
datasets are closer to a linear magnitude system than the results from 26, the
sense of the difference is that faint stars ($15 < V < 21$) have in general been
measured too bright compared to brighter ones on those nights when \ngc{2419}\
was observed.  This is in the sense found by Saha \etal:  that I measure faint
stars in \ngc{2419} brighter than they do.  Dividing the magnitude differences
into two bins, respectively brighter and fainter than $V = 15$, and counting
only those stars with acceptable measurements in both $V$ and $I$ in both
datasets, in the brighter bin the median magnitude difference in $I$ is
--0.002$\,$mag (336 stars) and in the fainter bin it is +0.005$\,$mag (2,720
stars).  In both bins the median difference is 0.000$\,$mag in $V$, for the
same numbers of stars.

If this effect is real, it {\it could\/} therefore account for an additional
0.007$\,$mag portion of the offset noted by Saha \etal, except for the fact that
the 26 with-2419 datasets having $I$-band data include 13 from the WIYN 3.5m and
0.9m telescopes.  Figs.~8 and 9 indicate that the WIYN data participate in the
bright/faint zero-point difference:  in the WIYN data the median magnitude
difference brighter than $V=15$ is --0.001$\,$ mag, and fainter than $V=15$ it
is +0.011$\,$mag.  In comparison, for the non-WIYN datasets that include
\ngc{2419} observations, the corresponding median differences are --0.006$\,$mag
and +0.003$\,$mag.  If the $I$-band nonlinearity is present in the WIYN
instrumentation or observing conditions, it should affect the analysis performed
by Saha \etal\ as well as mine.  If, on the other hand, the nonlinearity is an
artifact of my data processing, it is hard to see why it would affect primarily
the $I$ band on those nights when \ngc{2419} was observed, and not the other
filters on those same nights or the $I$ filter on other nights.  A similar
effect is {\it not\/} seen in the aggregate data for $B$, $V$, or $R$ (Fig.~7),
which reduces the likelihood that the bright-faint nonlinearity in $I$ is the
result of a defect common to several different cameras, or a minor bug in 
the reduction software.  Perhaps it is a fluke of finite-number statistics and
should be ignored.  In fact, these data show very little evidence for a simple
nonlinearity of the detector response from the \ngc{2419} nights as compared to
the non-2419 nights: such nonlinearity would normally be expected to grow
exponentially toward bright magnitudes since at faint magnitude levels the small
flux difference between the (star+sky) measurement and the (sky) measurement
offers little lever arm for a nonlinearity to act upon.  

Fig.~10 shows the median magnitude differences between the non-2419 nights
and the \ngc{2419} nights binned by \vmi\ color.  Again I have considered only
stars with at least four observations under photometric conditions and standard
errors of the mean magnitude not greater than 0.03$\,$mag in {\it both\/} of the
two samples, as well as no evidence of intrinsic variability in excess of
0.05$\,$mag, r.m.s., when all observations are considered together.  Apart from
the very reddest stars, it is evident that the color terms in the
transformations are as well controlled on the \ngc{2419} nights as they are on
the non-\ngc{2419} nights.  Anent the reddest stars, Landolt himself lists only
six stars with \vmi\ colors redder than 3.50, and one of these---98~L5---has
claimed uncertainties $>\,0.10\,$mag in both $V$ and \vmi, while
another---92~427---was observed only twice on a single night.  These two
therefore hardly count as standard stars.  Of the other four very red Landolt
standards, G3-33, G12-43, G156-31, and G45-20, only the last has 
observations in my database: four CCD images consisting of one in $V$ and one in
$I$ from each of two different observing runs, one of which was photometric and
the other of which was not, and neither of which included \ngc{2419}.  Therefore
all the stars redder than $\vmi \approx 3$ that are plotted here or discussed
elsewhere in this paper rely heavily upon extrapolation of the color
transformations.  Consequently it is not surprising that their absolute colors
are somewhat vaguely defined.  For completeness, I should note that the reddest
three bins for the $R$ bandpass contain, respectively, two, one, and one stars;
the large scatter seen here is therefore not a major cause for concern.  Another
thing to note is that the $I$-band magnitudes of very blue stars are
particularly hard to define consistently, as indicated by the large dispersion
(error bars) in the bluest few bins.  This effect has been seen before (\eg,
Stetson, Bruntt \& Grundahl 2003, Fig.~3).

\subsection{Are there any problems specific to the WIYN datasets?}

The remainder of this section will concern itself primarily with the data
from the WIYN 3.5m telescope, because the WIYN 0.9m data carry relatively little
of the weight in determining the mean photometric indices for the faint
\ngc{2419} standards.

\bigskip

\centerline{5.3.1.  Magnitude/exposure-time effect?}

When I carefully examined the photometric residuals from the CCDSTD reduction of
the primary and secondary standard stars observed with the WIYN 3.5m
telescope, I noticed an offset between the residuals of the bright---primary and
secondary standards in a ratio of about 1:15---and faint---entirely
secondary---standards.  For example, Fig.~11 shows $\Delta V$ {\it versus\/} $V$
(left-hand panels) and $\Delta I$ {\it versus\/} $I$ (right) for chip~1 (top
panels) and chip~2 (bottom) of MiniMos from the initial reductions of the
``wiyn3'' datasets from the 3.5m telescope.  The sense of the differences is
observed instrumental magnitude as measured from the CCD image {\it minus\/} the
instrumental magnitude that is predicted by Eq.~1 above, using the known
standard-system magnitudes of the stars and the best-fitting values of the
color-transformation and extinction coefficients.  In each case, the fitting
residuals appear to define two distinct clumps, each one flaring out with
larger dispersions toward fainter magnitudes.  However, in each panel the clump
that contains the majority of the apparently bright photometric standards is
systematically shifted toward negative numbers:  these stars' instrumental
magnitudes have been measured too small (too bright) compared to most of the
fainter standards.  Within each clump, there is no obvious tendency for the
centroid of the magnitude residuals to vary systematically with apparent
magnitude; rather the appearance is that the brighter standards have been
translated in bulk toward negative fitting residuals.

In the ``wiyn3'' observing run, this effect appears to be roughly consistent for
both chips and both filters, but it does not appear to be strictly constant from
run to run:  for instance, Fig.~12 shows the corresponding residuals from
initial reductions of the ``wiyn4'' run.  It is not obvious to the eye, but when
actual weighted averages are calculated in this dataset the effect is found to
be in the opposite sense to that observed in ``wiyn3'', with the brighter stars
having more positive residuals in ``wiyn4''; a hint of this is perceptible in
the bottom right panel.  A similar effect is {\it not\/} seen in most of the
other datasets that include observations of \ngc{2419}:  for instance, the
three-night observing run that I have labeled ``abi36,'' carried out by Abi~Saha
on the WIYN 0.9m telescope, shows no obvious systematic offset between the
bright standards and the faint ones: Fig.~13 illustrates 8,240 fitting residuals
in $V$ and 9,580 fitting residuals in $I$ from all observations of primary and
secondary standards on the three nights.  Similar plots could be produced from
the remaining datasets.  The variability of the bright/faint offset from run to
run and from telescope to telescope suggests that this is not a general
systematic problem within the secondary standard system itself.  

Figs.~14 and 15 demonstrate that the problem could be viewed as relating to
exposure time rather than to apparent magnitude.  Again, residuals in $V$ are
shown in the left panels and $I$ on the right, and chip~1 is at the top and
chip~2 is at the bottom.  This time, the abscissa is exposure time displayed
on a logarithmic scale.  Each short, thin horizontal line represents the
photometric residual from a single observation of a primary or secondary
standard star.  These figures show the same stars as are illustrated in Figs.~11
and 12, but since the exposure times are quantized here there is a much greater
overlap of points than in the previous figures.  A longer, heavier line
shows the value of the median photometric residual at each distinct exposure
time.  In every panel for the ``wiyn3'' run, the shortest exposures are
associated with the most negative residuals, on average.  The longer exposure
times, especially those represented by many different standard-star
observations, have more consistent fitting residuals.  For the ``wiyn4'' run
(Fig.~15), again the trend is not obvious in the $V$ band data, where the
shortest exposure time was 6$\,$s, but it is perceptible in the $I$-band data,
where the shortest exposure was 3$\,$s.  And again, when suitable weighted
averages are taken, it is found that overall the shortest exposures in the
``wiyn4'' datasets have more positive residuals than the longer exposures, on
average.  

I found that these trends could be removed to first order by assuming that the
exposure times were not exactly the values given in the image headers.  For
instance, if I assume that for the run ``wiyn3'' all exposures are 0.06$\,$s
longer than the integer number of seconds given in the image headers, I could
remove the net difference between the median residual of the shortest exposures
and the median residual from all exposures of 20$\,$s or greater, when data from
both chips and all filters are considered simultaneously.  In the case of the
``wiyn4'' data, the offset was removed by assuming that the exposures were
0.05$\,$s {\it shorter\/} than indicated in the image headers.  For the
``wiyn2'' dataset the estimated adjustment was +0.02$\,$s; for the observing
that I have named run ``wiynb''---which included no observations of
\ngc{2419}---the correction was +0.04$\,$s; and for ``wiync'' it was 
+0.02$\,$s.  The effect thus ranges from stars in the shortest exposures being
measured 0.02$\,$mag too faint (the shortest ``wiyn4'' exposures being
2.95$\,$s rather than 3$\,$s) to 0.03$\,$mag too bright (2.06$\,$s {\it
versus\/} 2$\,$s in ``wiyn3'') when integer-second integrations are assumed. 
Exposures longer than expected appear to dominate over exposures that were
shorter than expected, contributing to a net stretch of the magnitude scale
between bright stars and faint in the WIYN 3.5m data.  This can lead to a
systematic offset in the \vmi\ colors, since the $I$-band exposures were
often shorter than those in $V$.  The effect observed here is in the same
sense as the $I$-magnitude offset found by Saha \etal:  according to them, my
$I$-band magnitudes for \ngc{2419} stars were too bright by about 0.04$\,$mag,
\ie, the flux ratio between my measurements for the bright standard stars and
the faint cluster stars was too small, compared to theirs.  

The net effect of altering the exposure times, when averaged over all exposure
times and applied to all primary- and secondary-standard star observations, has
little influence on my photometric results:  when the calibrated photometry from
all the WIYN 3.5m observing runs is combined, and adjusting only the photometric
zero-points without altering the estimated color or extinction terms,  the net
effect of going from integer- to fractional-second exposure times is to brighten
my $V$ results for \ngc{2419}\ by 1.3$\,$mmag and $I$ by 1.4$\,$mmag (and
1.6$\,$mmag in $B$ and 0.7$\,$mmag in $R$).  The smallness of these changes is
due to the predominant weight carried by the large number of secondary standard
stars that were measured in comparatively long-exposure images.  Results based
upon the fractional-second exposure times are what I have discussed in the
previous section of this paper.

However, Saha \etal\ did not use the secondary standards available in the 
Landolt fields or the program fields that they observed:  they based their
calibrations on only Landolt (1992) standards---all observed in short-exposure
images---and applied those calibrations to program objects in long-exposure
images.  To estimate the size of the exposure time effect under these
circumstances, I reperformed the calibration of the WIYN 3.5m photometric data. 
This time I employed only the Landolt (1992) standards to derive the
transformations, and again I enforced the color transformations and extinction
slopes that I had derived from the full set of primary and secondary standard
observations, leaving only the zero-points to be determined by the software.  In
this case, going from the integer-second integration times to the
fractional-second integrations caused the calibrated photometry for \ngc{2419}
to brighten by 0.015$\,$mag in $V$ and 0.017$\,$mag in $I$.  This accounts for
most of the 0.021$\,$mag discrepancy that remained between my results and those
of Saha \etal\ after correcting the color-slope error of my results posted in
September 2004.  

None of this demonstrates that there was an actual problem with the timing of
the camera shutter on the WIYN telescope.  In fact, Saha \etal\ direct their
readers' attention to tests indicating that the MiniMos shutter is accurate to
0.2\% for exposures as short as 0.3$\,$s.  This implies an absolute accuracy of
600$\,\mu$sec in the shutter mechanism after correcting for known software
delays.  (The document cited itself says only, ``Over the entire effective range
of exposure times from 0.2$\,$s to 10$\,$s, no effective non-linearity is seen
at the one percent level,'' on the basis of experiments conducted with the
flat-field lamp.  It is not stated whether these tests were carried out on more
than one occasion, at more than one temperature, with the telescope in various
positions, or with the shutter being exercised after varied periods of
inactivity.)  Fudging the exposure times is merely a numerical kludge that
empirically removes an observed trend, much as a low-order polynomial in some
color index is presumed to calibrate away filter-bandpass mismatch.  Nor can I
argue that a single alteration to the integration time is appropriate for all
exposures from a given  night of observations.  This merely appears to remove
the net trend on average; any variation in whatever is causing the effect
contributes to the random errors remaining in the results.  Saha \etal\ would
have had great difficulty in detecting this effect, whatever it is, since they
based their calibrations solely on bright standards in short-exposure images.  

\bigskip

\centerline{5.3.2.  Color effects?}

In reducing the WIYN 3.5m data, I found strong evidence for trends in the
calibration that were quadratic in the color term.  Specifically, the top panel
of Fig.~16 contains the residual $I$-band magnitude differences between my
current adopted magnitudes and the average results derived from the WIYN 3.5m
data alone for all primary and secondary standards observed during those
observing runs (including ``wiynb'', which provided no observations of
\ngc{2419}) when the best linear calibration is employed: $$i \sim I - 0.050(\pm
0.002)(\vmi).$$ Here I have plotted residuals for 1,533 distinct primary and
secondary standard stars having at least four observations and $\sigma_{\bar I}
\leq 0.03\,$mag in the WIYN 3.5m datasets.  This may be compared to the Saha
\etal\ calibration for one of the chips (not specified) from the run I have
named ``wiyn2'': $$ I \sim i - 0.024(\hbox{\it v--i\/})$$ based upon 22
observations of Landolt (1992) standards.  Saha \etal\ obtained an independent
color transformation for each chip on each night.  Fig.~3 above indicates that
my $I$-band magnitudes are now on the Landolt system for $-0.6 < \bmi < +5.5$,
which corresponds to $-0.3 < \vmi < 2.9$  (the effective limits of the color
range well sampled by Landolt standards).  Thus, the extreme curvature seen at
the red end is largely inferred from secondary standards whose colors have been
extrapolated beyond Landolt's limits.  However, these data represent the average
of many different extrapolations with many different CCDs and filters, not all
of which required quadratic color terms.  Even within the range spanned by the
Landolt standards, the offsets seen in the upper panel of Fig.~16 at $\vmi <
0.5$ and $2.0 < \vmi < 3.0$ are real and statistically highly significant.  

The lower panel shows the corresponding residuals for the posted standard stars
from my best-fitting quadratic calibration:
$$i \sim I - 0.0106(\pm 0.0012)(\vmi) - 0.0133(\pm 0.0012)(\vmi)^2.$$
Similarly, my best-fitting transformation curve for all the other filters in 
the WIYN 3.5m data were 
$$b\sim B+0.028(\pm 0.008)(\bmv)-0.033(\pm 0.013)(\bmv)^2,$$
$$v\sim V-0.010(\pm 0.002)(\vmi)+0.001(\pm 0.001)(\vmi)^2\hbox{, and}$$
$$r\sim R+0.043(\pm 0.005)(\vmr)-0.053(\pm 0.029)(\vmr)^2,$$
which may be compared to the relations in Saha \etal, Eqs.~5--8.
The chip-to-chip and run-to-run differences in these coefficients were small, $<
\pm 0.004$ in a root-mean-square sense, so I concluded it was adequate to
assign a common color transformation to all the datasets.  If this
simplification is invalid, it means that for some datasets the color slope has
been underestimated while for others it has been overestimated.  In this case,
in some datasets the reddest stars will be calibrated too bright and in others
they will be calibrated too faint.  The net deleterious effect of the
oversimplification will largely average out for stars measured in many different
observing runs so---even if it is unjustified---it is not likely to cause major
systematic errors in the final standard list, especially for stars of neutral
color.

For comparison, Fig.~17 shows the corresponding residuals for only the Landolt
(1992) standards when the average linear and quadratic fits are imposed on all
the WIYN 3.5m observing runs.  (These linear and quadratic fits are the same
ones discussed in the previous two paragraphs, and were performed on the basis
of all current primary and secondary standards, not on the basis of Landolt
standards only.  However, here I have plotted the fitting residuals in the sense
Landolt's published photometry minus the values that I obtain from my data only
for stars in common.)  Here I have placed no restrictions on the number of
observations or the size of the uncertainties, except that I have omitted the
extremely red star Landolt 98~L5, for which Landolt (1992) gives $V\,=\,$17.80
and \vmi$\,=\,$5.80 with large uncertainties, based on six observations per
filter from three nights.  My own observations (12 in $B$, 98 in $V$, 38 in $R$
and 79 in $I$) suggest that this star is a variable at a level of some
0.23$\,$mag (r.m.s.).  Saha \etal\ would not have known this, but it is clear
that they omitted the star anyway (it is patently absent from their Fig.~12, for
instance), probably because of its large published uncertainties.  This plot
does include a number of other stars that I would not have used as primary
standards (Landolt's stars 98~102, 98~634, 98~642, 98~646, 98~652, 101~262,
104~244, 104~339, 104~l2, 110~362, 110~L1, and 113~L1), because Landolt himself
observed them fewer than five times each.  Without the eye of faith, it would be
hard to distinguish which of these fits is better.  However, a close look does
indicate that the half dozen bluest stars and the half dozen reddest stars are
skewed toward positive residuals in the top panel. 

\centerline{5.3.2.  Small-number effects?}

Note that, apart from Landolt star 98~L5, Fig.~17 contains the average
photometric results for every Landolt standard observed on either CCD of the
mosaic from all WIYN 3.5m observing runs combined.  The number of stars observed
on a given chip on a given night will be much smaller than indicated here.  The
question remains: given the circumstances, which is better, to impose mean
extinction coefficients and mean color transformations on the different chips
and observing runs, or to determine the extinction and color transformations
separately for each CCD on every night?  Saha \etal\ have illustrated their
photometric residuals for Landolt standards from all nights considered together
in their Figs.~13 and 14.  However, these plots do not give a good sense of the
number and distribution of residuals from a given night on a single chip.

In Fig.~18 I show my weighted mean fitting residuals for the $V$-band
standard-star images against airmass for each of the five photometric WIYN 3.5m
nights, including ``wiynb'' when \ngc{2419} was not observed.  To reduce
crowding, each point represents the weighted average fitting residual for all
primary and secondary standards contained in a single CCD image, and small
filled circles are for chip~1 in the mosaic camera, while large empty circles
are for chip 2.  As before, residuals are defined in the sense observed
instrumental magnitude as measured from the CCD image minus the value predicted
from Eq.~1.  Error bars represent the standard error of the mean residual.  The
number in the lower right corner of each panel indicates the total number of
individual standard-star measurements (chips~1 and 2 combined) that are
represented by the points.  Fig.~19 is the corresponding plot for the $I$-band
observations.  

In performing these reductions, I concluded that night ``wiyn2'' was perceptibly
clearer than the others:  for this night I derived a weighted $B$-band
extinction coefficient of 0.20$\,$mag~airmass${}^{-1}$, which scaled to 0.11 in
$V$, and 0.07 in $I$ according to the standard extinction ratios given above (no
$R$-band data were taken that night).  For the other four nights my estimated
site-mean extinction coefficients were judged to be adequate: 0.24 in $B$, 0.14
in $V$, and 0.09 in $I$.  The degree of validity of these judgments is indicated
by the figures.  

On the night named ``wiync,'' of all the observations in {\it BVRI\/} there
are five CCD images whose absolute mean residuals are greater than 0.1$\,$mag.
Fig.~20 shows, at a compressed scale, the mean residuals in all four filters
plotted as a function of the time of night.  Again, the error bars represent
the formal standard errors of the weighted mean residuals.  There are two bad
images in the $R$ filter that were taken just before 6:00 o'clock; these come
from a single exposure with chips~1 and 2 of the Landolt standard field Ru149,
and contain 194 and 59 primary and secondary standards, respectively (counting
only Landolt's primary standard stars, there are eight and zero in the two
chips).  There is a pair of bad images in the $V$ filter taken just before 10:00
o'clock resulting from a single exposure of the Pal~4 program field; these
images contain 64 and one secondary standards, respectively (there are no
primary standards in the Pal~4 field).  Finally, there is one bad $B$-band
image; this is a chip~1 image containing a single secondary standard from the
same visit to Pal~4 as the two bad $V$-band exposures.  The chip~2 image from
the same exposure is marginally discrepant, with a weighted mean residual of
$+0.045\pm0.005\,$mag based upon 70 secondary standards.  I have no way of
knowing why these particular images are anomalous.  The other 99 images in $B$,
$V$, $R$, and $I$ from both CCDs on that night are basically fine,  although
there may be some other subtle indications that night ``wiync'' might not have
been completely photometric: Fig.~20 shows some evidence of instability from the
beginning of the night until 6:00 or 6:30, and Figs.~18 and 19 show slightly
enhanced scatter in the ``wiync'' data compared to the other nights.  At this
point, the astronomer must make a binary decision, whether to ignore the
discrepant images and treat the rest of night ``wiync'' as photometric, or to
reduce  the night's data in non-photometric mode;  both Saha \etal\ and I have
taken the former approach.  

For comparison, in Figs.~21 and 22 I illustrate the frame-averaged fitting
residuals for the Landolt (1992) standards only.  For these reductions I
enforced my own estimated values of the quadratic color transformations and
extinction coefficients; I have left only the photometric zero-points free to be
determined by the fitting procedure.  Again, small filled circles represent
individual-image weighted mean residuals from chip~1, large empty circles
are for chip~2, and the errorbars represent the standard error of the mean
residual.  The numbers in the lower right corner of each panel represent
the number of individual photometric measurements shown in the plot:  if
a given star has been observed multiple times with a particular chip/filter
combination during the night, all the observations are included in the total
number count since it may have been observed at different airmasses.  It is
evident that {\it if\/} one chose to determine the extinction coefficients
completely freely from these data, it would sometimes be possible to derive
quite different values from the ones I have adopted based on the full set of
primary and secondary standards.  To the extent that the program fields were not
observed at precisely the same airmass, on average, as the Landolt standard
fields, any indeterminacy in the extinction coefficients could contribute to
systematic errors in the calibrated photometry for the science target fields
observed on those nights.  

Figs.~23, 24, and 25, and 26 address the question of whether it is adequate to
use the same color transformation for both chips and all observing runs.  In
them I have plotted, binned by \vmi\ color, the median residuals of the
standard-star observations (primary and secondary standards) separated by
observing run (five panels in each plot), and by chip and filter (separate
plots).  The data are the same as were used for Figs.~18 and 19.  Here, as in
Figs.~7--10, for instance, the error bars represent the $\pm 1 \sigma$ spread in
the residuals within each bin; points with no perceptible error bars usually
represent only one or two stars.  For instance, the two dots with $\vmi > 4.0$
for run ``wiyn4'' in all four figures represent only a single star in each case
(secondary standards L110-S96 with $\vmi = 4.98$ and L110-S44 with $\vmi =
5.27$).  The total number of different primary and secondary standards
represented in each panel is given in the lower right corner; a star with
multiple observations in a given filter/chip combination on a given night is
counted only once, since its true color is always the same.  One can imagine
that there are significant deviations from the zero line in some of the bins. 
Overall, however, it seems that there is no systematic slope in the
well-populated regime, $0.0 \leq \vmi \leq 2.5$ in any of the panels, so the
adopted constant color transformation appears to do no {\it serious\/} injustice
to the data from any one chip/filter/date combination. 

For comparison, Figs.~27, 28, 29, and 30 show the magnitude differences for only
the individual Landolt (1992) standards in the various datasets.  These figures
are based on are the same reductions as Figs.~23 through 26:  unlike the case
with Figs.~21 and 22, they are calibrated using my zero-points as well as my
estimates of the extinction and color coefficients.  However, here I have picked
out the final magnitude differences only for Landolt (1992) standards, including
those that have fewer than four Landolt observations or published standard
errors larger than 0.03~mag.  Since these plots are not crowded, I have not
plotted median residuals in color bins; each point represents the net magnitude
difference for a single Landolt star---the nightly average difference if the
star was observed more than once with a particular chip/filter combination.  The
number in the lower right corner of each panel indicates the number of different
Landolt stars used in the comparison.  It seems that if one were to base the
calibration of the data on only these stars and no others, it would be possible
in some cases to infer different zero-points and color-transformation slopes
than those that I have adopted from the larger sample of primary and secondary
standards.  

\section{SUMMARY}

In this paper I have attempted to find explanations for the different $I$-band
photometric calibrations between my posted photometric standard values and the
results reported by Saha \etal\ (2005).  In the course of this investigation I
have discovered and---I hope---largely corrected a systematic error in my
$I$-band magnitude system which depended on the temperature of the star.  This
systematic effect produced maximum errors of about $\pm\,0.02\,$mag for stars at
the extremes of Landolt's color range:  $\bmi \sim -0.6$ and $\bmi \sim +5.0$,
or $\vmi \sim -0.35$ and $\vmi \sim +2.65$, or $\bmv \sim -0.3$ and $\bmv \sim
+2.3$.  Systematic errors in my previously posted magnitudes probably were not
in excess of 0.005$\,$mag for the most common stars: those with $0.6 \ltsim \vmi
\ltsim 1.4$.  In this range the average systematic error is close to
0.000$\,$mag, and the root-mean-square contribution to photometric scatter is
probably $\sim\,0.003\,$mag; over the whole color range, the r.m.s.\ error is
slightly less than 0.008$\,$mag.  This mistake is probably responsible for about
20\% of the discrepancy noted by Saha \etal\ \  I have subsequently done my best
to eliminate this part of the anomaly.

In comparing 33 datasets that include observations of \ngc{2419} to the
photometric results from 649 datasets that do not include data for the cluster,
I conclude that the two subsamples are on the same photometric system to
$\sim\,\pm 0.003\,$mag, which is consistent with the notion that, even under
the best of circumstances, the average results for any one field from any one
night of observations---due to anomalous extinction, flat-fielding errors,
shutter-timing errors, or whatever---can be systematically incorrect by
$\sim\,0.02\,$mag, and that these systematic errors can be beaten down by
combining many independent nights' worth of data.  

My analysis of the WIYN data has differed from that of Saha \etal\ in four
significant particulars.  First, I employed my expanded standard list, which
consists of the average of my results from 179 observing runs with Landolt's
results for stars in common, plus an additional $\sim4,300$ secondary standards
that fell within the WIYN images; Saha \etal\ used $\sim110$ (depending upon the
filter) different stars from Landolt (1992).  Second, my analysis of
the 3.5m data initially produced a systematic difference in the photometric
residuals of the bright primary and secondary standards as compared to those of
the faint secondary standards.  This discrepancy could be removed to first order
by making empirical adjustments to the exposure times as recorded in the image
headers.  The same adjustments corrected the chip 1 and chip 2 residuals and the
$V$- and $I$-band residuals comparably well, but different adjustments were
indicated for different observing runs.  Third, my analysis used quadratic color
corrections to transform the WIYN instrumental magnitudes from both telescopes
to the standard photometric system, while Saha \etal\ found linear
transformations to be adequate.  These latter two effects, taken together,
account for most of the $\sim0.04\,$mag discrepancy found by Saha \etal\ \
Fourth, I estimated the color and extinction corrections from multiple datasets
(different chips, filters, nights) and imposed mutually consistent, constant
values for these coefficients on the final photometric reductions.

The necessity for altering the exposure times and adopting quadratic color
transformations resulted entirely from my use of the secondary photometric
standards in addition to the primary Landolt (1992) standards.  Both of these
effects are unseen when only the bright Landolt standards are employed.
The additional standards do not make the fourth alteration of the reduction
protocol {\it necessary\/}, but I believe that they demonstrate that the
homogenized extinction and color corrections are {\it adequate\/}. Short of
having Arlo Landolt observe a representative sample of my secondary standards, I
have no further proof that my results for the secondary standards are on the
Landolt system.  However, I have shown that my observations and my software can
place my observations of Landolt's stars on his system, and it is logical to
expect that the same lines of code and the same transformation formulae should
work equally well on other stars observed with the same equipment, provided that
the systematic errors occasioned by inherent differences between the 
observations of Landolt standards and the program stars are adequately
controlled.  To the extent that this latter statement is untrue---if a given CCD
is nonlinear, for instance, or if the standard and program stars are not
observed at similar times of night or at similar airmasses---one hopes that such
failures will average out over many nights with many cameras.  When all else
fails, more observations are generally better than fewer.

\acknowledgments

I would like to thank Abi~Saha for many stimulating discussions, and I'd like
to thank him again and the WIYN observatory staff in general for providing me
with their data from six observing runs.  I am very grateful to Michael~Bolte,
Howard~Bond, and Nicholas~Suntzeff, who also generously donated some of the data
that were used in this study.  I continue to be very grateful to the Canadian
Astronomy Data Centre and the Isaac Newton Group Archive for providing data for
this and other ongoing studies.  

\appendix

\section{Weighting published photometric indices}

To obtain maximum benefit from the combination of observations, it is
important to consider the uncertainties of the individual measurements and
take them into accout by weighting the data appropriately.  Assume that
a set of measurements, $x_i$, for $i=1,\ldots,n$, are made of of some physical
quantity, $x$, with measuring errors that are known to possess accurately
Gaussian probability distributions with known standard deviations $\sigma_i$. 
It is an elementary exercise to demonstrate that a maximum-likelihood estimate,
$\bar x$, of the unknown true value of $x$ is achieved by assigning a weight
$w_i \propto \sigma_i^{-2}$ to each individual observation.  In the still more
simplified case where all the $\sigma_i = \sigma$---a known, constant
value---$\bar x = (\sum x_i)/n$, and a $\pm\,1\sigma_{\bar x}  =$ 68\% confidence
interval for the average measurement is given by 
$$\sigma^2_{\bar x} = \sigma^2/n. \eqno{(A1)}$$ 
It is similarly elementary, but a different problem, to show that when the
standard deviation of the measuring errors, $\sigma$, is unknown, but is
presumed to be the same for all $i$, an unbiased estimate of the true standard
deviation can be derived from
$$\bar\sigma^2 = {{\sum_{i=1}^n (x_i - \bar x)^2\over{n-1}}}. \eqno{(A2)}$$

It is commonly seen in the astronomical literature that the value of the
quantity being sought and the standard deviation of the measuring error
are {\it both\/} considered to be unknown but constant.  In this
case the equal-weighted mean quantity $\bar x$ is generally calculated as above,
the standard deviation of one measurement error is estimated from Eq.~(A2), 
the resulting value of $\bar\sigma$ is substituted for $\sigma$ in Eq.~(A1), and
the value of $\sigma_{\bar x}$ given by 
$$\sigma^2_{\bar x} = {{\sum_{i=1}^n (x_i - \bar x)^2\over{n(n-1)}}} 
\eqno{(A3)}$$
is published as the $\pm 1\sigma$ = 68\% confidence interval for the true value
of $x$ about $\bar x$.  This procedure is not strictly legitimate, and is
significantly wrong for small values of $n$.  

Consider this:  if exactly two measurements of some unknown quantity
$x$ are obtained, what is the numerical probability that the true value
of $x$ lies {\it between\/} the two measured values, $x_1$ and $x_2$?
To clarify the discussion, and without loss of generality, we may suppose
that $x_1 = 1$ and $x_2 = -1$.  Then $\bar x = 0$, $\bar \sigma^2 = \left[1^2 +
(-1)^2\right]/(2 - 1) = 2$, and $\sigma^2_{\bar x} = \bar\sigma^2/2 = 1$.
Therefore $0\pm1$ defines a $\pm 1\sigma$ = 68\% confidence interval for the
true value of $x$.  This is not the right answer.

The first of the two measurements is certain to be either greater or smaller
than the true value of $x$; provided the measuring error is not identically
zero, the probability of measuring exactly the right value is nil and 
an overestimate is just as likely as an underestimate.  The
second measurement, provided it is statistically independent of the first,
also has equal likelihood of being too large or too small.  The probability
that the second measurement lies on the same side of the true value as
the first measurement is precisely 50\%, and the probability that the two 
measurements straddle the true value is also 50\%.  In the example given above,
the interval $0\pm1$ therefore represents a 50\% confidence interval, not 68\%.

The problem is that the true value of {\it neither\/} $x$ nor $\sigma$ is 
known.  Both are estimates based upon uncertain data, and when a confidence
interval for a result is claimed, it must take into account the fact that {\it
both\/} $\bar x$ and $\bar\sigma$ are {\it estimates\/} based on uncertain data.
The complication arises from the fact that, while the likelihood distribution
for the true value of $x$ is symmetric about the estimate $\bar x$, the
likelihood for the true value of $\sigma$ about $\bar\sigma$ is not:  if
the estimate is $\bar\sigma = 1$ it is absolutely impossible that the true value
of $\sigma$ could be 0, but it is quite plausible that the true value could be
2.  It is even possible---though unlikely---that the true value of $\sigma$
could be 10.  Therefore, when both $\bar x$ and $\bar\sigma$ are estimated
from the same small dataset, the possibility that the measuring errors have been
greatly  overestimated is slight, but the possibility that they have been
underestimated is not negligible.  

The confidence interval for the true value of the quantity being sought should
be estimated from Student's $t$-distribution, not from the Gaussian
distribution.  When $n=2$ a 68\% confidence interval is equal to $\pm
2\bar\sigma$.  Many years ago I did a little experimenting with a calculator and
convinced myself that the true range of a 68\% confidence interval based on
Student's $t$-distribution is quite well approximated by 
$$\sigma_{\bar x} = {(n-1)\over(n-1.5)} {\bar\sigma\over\sqrt{n}}$$
This is accurate at $n=2$ and completely correct as $n\rightarrow\infty$, and is
wrong by no more than a few percent for intermediate values of $n$.  Since
Landolt's published uncertainties appear to be based on Eq.~A3, this correction
results in an increase of his error estimates by about 14\% when $n=5$, which is
the smallest number of observations I am willing to accept for a fundamental
standard star.  When I average Landolt's measurements of a star which he has
observed twice with three or more of my own measurements to define a standard
star with at least five independent observations, I double his published error
estimate in defining the relative weights.  The errors of my own photometry are
based on readout noise, photon statistics, and the quality of the profile fits
{\it as well as\/} on observation-to-observation agreement.  Therefore the
correction factor that I apply is close to unity if the
observation-to-observation agreement is consistent with the other error
estimates.  If the mean error of unit weight is significantly greater than
unity, then I apply a corresponding fraction of the Gauss-to-Student correction.

After the above scaling of the standard errors, I also apply an further additive
standard error of (0.001$\,$mag$)/\sqrt{12}$, in quadrature, to the error
estimates to allow for the error induced by Landolt's having rounded off his
published photometric indices to three decimal places.  This is a non-negligible
effect when the published uncertainty is $\ltsim0.001\,$mag, as is often the
case for Landolt standards with more than ten or twenty observations.  

Finally, for a fair combination of Landolt's results with my own, I must deal
with the fact that he works in colors and I work in magnitudes.  That is,
Landolt publishes indices and standard errors in $V$, \bmv, \vmr, and \vmi, for
instance, whereas my software operates on $B$, $V$, $R$, and $I$ and their
associated standard errors.  The reason for the different approach lies in the
equipment.  With a photomultiplier, it is easy to cycle rapidly and
repeatedly through the various filters.  If the observing sequence is symmetric,
the aggregate observations in the different bandpasses all occur at the same
effective mean airmass, and temporal variations in the atmosphere and equipment
cancel out in the flux ratios among the various filters as long as they are slow
compared to a single integration cycle.  Instrumental colors are therefore well
defined.  With CCD observations, the assumption that the different filters are
observed at the same effective airmass and time is harder to justify, and the
definition of instrumental colors is therefore problematic.  In addition,
farther down the reduction pipeline, the efficient combination of results from
different nights is more reliable when magnitudes are the coin of the realm:
as a particular case in point, CCDAVE is able to employ the mean of all
observations of a given star from {\it all\/} datasets in defining the
standard-system colors to use in the transformation equations for {\it each\/}
dataset.  When transformations are based on instrumental colors, the color
correction appropriate to a given observation is determined from a single pair
of error-prone measurements that must be closely matched in time and airmass.

Adopting magnitudes as the fundamental photometric indices necessitates
converting Landolt's standard errors in color to standard errors in magnitude,
which cannot be done with full mathematical rigor.  For faint stars with
observations whose precision is limited by photon statistics, the errors in $B$
and \bmv, for instance, will be correlated, but the errors in $B$ and $V$ will
be uncorrelated.  In this case, $\sigma^2(\bmv) = \sigma^2(B) + \sigma^2(V).$ At
the other extreme, the $B$ and $V$ magnitudes might have their errors dominated
by extinction fluctuations, in which case those errors would be highly
correlated, while those same fluctuations could cancel out in \bmv\ due to the
rapid cycling through the filters.  If the errors in magnitude and color are
completely uncorrelated, then one would expect that $\sigma^2(B) = \sigma^2(V) +
\sigma^2(\bmv)$.  The usual case will be somewhere between these limits, 
$$\hbox{Max}\left[0, \sigma^2(\bmv) - \sigma^2(V)\right]\leq \sigma^2(B) \leq 
\sigma^2(V) +\sigma^2(\bmv),$$
and without knowing the correlation coefficient between $B$ and $V$,
not much more can be said.  Therefore, I use
$$\sigma^2(B) = {1\over2}\left[\sigma^2(V) + \sigma^2(\bmv)\right]$$
as my estimate of the standard error of Landolt's  mean $B$ magnitude,
with equivalent definitions  for the other filters.

\clearpage

\figcaption[fg1.eps]{This figure shows the differences
between Landolt's (1973, 1983, 1992) published photometry and the final
results (as of September 2004) of my analysis of 594 photometric datasets.  I
have plotted magnitude differences only for stars that were observed at least
four times by Landolt and had at least four measurements under photometric
conditions in my data, had standard errors of the mean magnitude no larger than
0.03$\,$mag in both samples, and showed no evidence of intrinsic variation
larger than 0.05$\,$mag, r.m.s., in my data.  Each individual Landolt standard
star meeting these criteria is represented by an error bar indicating the
$\pm1\sigma$ confidence interval obtained by combining Landolt's standard error
of the mean with mine, in quadrature.  Data are plotted for 230 stars in
$B$, 261 in $V$, 144 in $R$, and 163 in $I$.}

\figcaption[fg2.eps]{The individual magnitude differences illustrated
in Fig.~1 have been binned in 0.5-mag intervals of \bmi\ color, and the
median magnitude difference in each bin is designated by a filled circle.
The error bars represent $\pm\sqrt{\pi/2}$ times the mean absolute deviation
from the median, which is a robust measure of the dispersion of the
differences.  The $I$-band data show a clear trend of $\left<\Delta I\right>$
with color, in the sense that redder stars have more positive values of
$I$(Landolt)--$I$(Stetson).  A straight line fitted to the weighted magnitude
differences in $I$ is shown as a dashed curve.}

\figcaption[fg3.eps]{As in Fig.~2, this figure shows the median 
value and the dispersion of the magnitude differences between Landolt's
published results and mine for stars meeting the acceptance criteria,
{\it after\/} my {\it BVRI\/} magnitude systems have been twisted to
match Landolt's as described in the text.  These are the data that were
posted to the World-Wide Web in January 2005.  The bin $5.0 < \bmi < 5.5$
contains only one star.  Some differences between Landolt's magnitude
systems and mine as a function of color remain, but they are not well
described by low-order polynomials in color.}

\figcaption[fg4.eps]{This figure illustrates the difference between
my calibrated photometry for \ngc{2419}\ stars as of September 2004---when
my posted data still suffered from the systematic color error illustrated
in Fig.~2---and my calibrated photometry as of January 2005, when the
systematic error had been reduced to the level shown in Fig.~3.  Also,
the January 2005 analysis is based on more datasets and more secondary
standards than the September 2004 analysis.  The upper panel shows
$\Delta V$ in the sense 2005 {\it minus\/} 2004 plotted against \vmi\ color,
and the lower panel shows $\Delta I$ versus \vmi.  
The error bars represent
approximate $\pm 1$ standard error of the mean magnitudes at {\it either\/}
epoch:  that is, the September 2004 standard errors and the January 2005 errors
have been added in quadrature and divided by $\sqrt2$.} 

\figcaption[fg5.eps]{This figure shows the difference between the
published photometry of Saha \etal\ (2005) and my results according to the
September 2004 calibration for \ngc{2419}\ stars meeting the acceptance
criteria listed in the text.  The upper panel shows $\Delta V$ in the
sense Stetson {\it minus\/} Saha plotted against the average of our two
estimates of \vmi\ color.  The bottom panel shows $\Delta I$ versus
\vmi.  The error bars are the result of adding Saha's estimated standard
errors in quadrature with mine, which probably slightly overestimates their
size.}

\figcaption[fg6.eps]{This figure is the same as Fig.~5, except
that I have compared the published photometry of Saha \etal\ to my results
according to the January 2005 calibration, after I have largely removed the
color-dependent calibration error in my magnitude scales.}

\figcaption[fg7.eps]{This figure shows a comparison between the
average photometric results for primary and secondary standard stars
contained in 616 datasets that do {\it not\/} contain any observations
of the cluster \ngc{2419}, and the average results from the 33 datasets
that {\it do\/} include data for \ngc{2419}.  This comparison does not
include the secondary standard sequence in \ngc{2419} itself, since
by design the first of these two subsamples contains no data for the cluster.
The individual stars have been binned into 1.0-mag intervals of $V$ magnitude;
the filled circles mark the median magnitude difference in each bin,
and the error bars represent a robust estimate of the $\pm1\sigma$ dispersion
among the differences in each bin.}

\figcaption[fg8.eps]{As in Fig.~7, except that here I have compared
the average results for primary and secondary standard stars from {\it only\/}
the 13 WIYN datasets that include \ngc{2419} data to the average results from my
616 datasets with no \ngc{2419} observations.  As before, the filled circles
represent the median magnitude differences within 1.0-mag bins, and the error
bars show a robust estimate of the $\pm1\sigma$ dispersions of the differences
within the bins.} 

\figcaption[fg9.eps]{As in Figs.~7 and 8, except that here I have
compared the average results for primary and secondary standard stars from {\it
only\/} the 20 non-WIYN datasets that include \ngc{2419} data to the average
results from my 616 datasets with no \ngc{2419} observations.  As before, the
filled circles represent the median magnitude differences within 1.0-mag bins,
and the error bars show a robust estimate of the $\pm1\sigma$ dispersions of the
differences within the bins.} 

\figcaption[fg10.eps]{This figure shows the same data as Fig.~7,
except that in this case the data are binned in 0.5-mag bins of \vmi\ color,
rather than in bins of $V$ magnitude.  Over the color range well sampled
by Landolt's primary standard stars, $\vmi < 3.5$, there is little evidence
for a systematic difference in the quality of the color transformations
from those 33 datasets that include \ngc{2419} data as compared to the
616 other datasets employed in my analysis.}

\figcaption[fg11.eps]{Magnitude residuals $\delta v$ (left panels)
and $\delta i$ (right panels), in the sense observed instrumental magnitude
{\it minus\/} the value predicted from the known standard magnitudes and the
transformation and extinction coefficients, plotted against the standard
$V$ and $I$ magnitudes, respectively, for MiniMos chip 1 (top) and 2 (bottom).
Each plotted symbol represents the residual obtained from a single observation
of a single primary or secondary standard star on the night that I have named
``wiyn3''.  The reductions have assumed that the integration times listed in the
FITS headers of the individual CCD images, which are always integer numbers of
seconds, are correct.  The distribution of fitting residuals breaks up into
two clumps.  In each of the clumps, the error distribution flares out toward
fainter magnitudes, but in the case of each chip/filter combination, the
clump consisting of the apparently brighter stars is systematically displaced
toward negative values relative to the fainter clump.  This indicates that the
measured instrumental magnitudes of the brighter standards have systematically
been measured too small (too bright) compared to the fainter standards.
}

\figcaption[fg12.eps]{Magnitude residuals $\delta v$ (left panels) and
$\delta i$ (right panels), in the sense observed instrumental magnitude {\it
minus\/} the value predicted from the known standard magnitudes and the
transformation and extinction coefficients, plotted against the standard $V$ and
$I$ magnitudes, respectively, for MiniMos chip 1 (top) and 2 (bottom).  As in
Fig.~11, each plotted symbol represents the residual obtained from a single
observation of a single primary or secondary standard star except that these are
now for the night that I have named ``wiyn4.''  In this case, the brighter
standards have not been shifted toward negative fitting residuals compared to
the fainter ones.  In fact, a quantitative statistical analysis indicates that
in this case the apparently brighter standards have been systematically shifted
toward positive residuals:  they have been measured too faint compared to the
fainter standards.  A hint of this effect is visible in the bottom right panel.}

\figcaption[fg13.eps]{Magnitude residuals $\delta v$ (top panel) and
$\delta i$ (bottom panel), in the sense observed instrumental magnitude {\it
minus\/} the value predicted from the known standard magnitudes and the
transformation and extinction coefficients, plotted against the standard $V$ and
$I$ magnitudes, respectively, for data obtained with the CCD ``s2kb'' on the
three-night observing run that I have named ``abi36.''  As in Figs.~11 and 12,
each plotted symbol represents the residual obtained from a single observation
of a single primary or secondary standard star.  Here, the data from all three
nights have been included in the same plot.  There is no obvious evidence for
a systematic bright-versus-faint split in the fitting residuals.}

\figcaption[fg14.eps]{The same photometric fitting residuals as are
illustrated in Fig.~11 are plotted here against integration time shown on a
logarithmic scale. As in Fig.~11, data from the night ``wiyn3'' are shown
with the $V$-band data on the left, the $I$-band data on the right, the
chip~1 data on the top, and the chip~2 data on the bottom.  Here the fitting
residual of a single observation of a single primary or secondary standard
star is shown as a short, thin horizontal line.  A long, thick horizontal
line marks the position of the median residual at each quantized
integration time.}

\figcaption[fg15.eps]{The same photometric fitting residuals as are
illustrated in Fig.~12 are plotted here against integration time shown on a
logarithmic scale. As in Fig.~12, data from the night ``wiyn4'' are shown
with the $V$-band data on the left, the $I$-band data on the right, the
chip~1 data on the top, and the chip~2 data on the bottom.  Here the fitting
residual of a single observation of a single primary or secondary standard
star is shown as a short, thin horizontal line.  A long, thick horizontal
line marks the position of the median residual at each quantized
integration time.}

\figcaption[fg16.eps]{These panels show the differences in $I$-band
magnitude between the weighted average calibrated photometry from all the
WIYN~3.5m observing runs taken together on the one hand, and my adopted
standard-system values (as of January 2005) on the other, plotted against
\vmi\ color.  Each errorbar represents the $\pm1\sigma$ standard error of the
mean difference based on all the WIYN 3.5m data.  The upper panel represents the
differences when the best-fitting linear color transformation against \vmi\
color is employed; the bottom panel is for the best-fitting transformation
employing a quadratic polynomial in \vmi.} 

\figcaption[fg17.eps]{This plot is based upon precisely the same data
and the same reductions as Fig.~16, except that here I have picked out the
$I$-band magnitude differences between the average WIYN 3.5m results and
Landolt's (1992) published magnitudes for the Landolt stars that were observed
during the course of the various WIYN 3.5m observing runs.  This plot also
includes a few Landolt (1992) standards with published magnitudes based upon
fewer than five observations, which were omitted from Fig.~16 because I impose a
minimum of five observations on stars which I accept as photometric standards. 
Landolt star 98~L5 has been omitted from this plot even though it was included
among the WIYN observations because its standard indices are extremely
uncertain.} 

\figcaption[fg18.eps]{In this figure, the weighted mean $V$-band fitting
residual of all the primary and secondary standard stars observed within an
individual CCD image has been plotted against the airmass at which the
observation was obtained, for those nights of WIYN 3.5m observations that
are considered photometric.  Small filled circles represent results for chip~1
of MiniMos, and large empty circles are for chip~2.  Error bars ($\pm1\sigma$)
are also plotted at the position of each point, but in nearly all cases they are
smaller than the point itself.  For WIYN 3.5m nights ``wiyn3,'' ``wiyn4,''
``wiynb,'' and ``wiync'' (bottom four panels) I used my adopted mean extinction
coefficients for the Kitt Peak site; on night ``wiyn2'' I estimated that the
extinction was roughly 20\% smaller than the site mean values.  The total 
number of distinct observations of primary and secondary standards made in the
$V$ filter on each night (in chips 1 and 2 combined) is indicated in the bottom
right corner of each panel.} 

\figcaption[fg19.eps]{The weighted mean $I$-band fitting residual of all
the primary and secondary standard stars observed within an individual CCD image
has been plotted against the airmass at which the observation was obtained.  As
in Fig.~18, data are plotted for the photometric WIYN 3.5m nights, where small
filled circles represent results for chip~1 of MiniMos, and large empty circles
are for chip~2.  Error bars ($\pm1\sigma$) are also plotted at the position of
each point, but in nearly all cases they are smaller than the point itself.  The
total number of distinct observations of primary and secondary standards made
in the $I$ filter on each night (in chips 1 and 2 combined) is indicated in the
bottom right corner of each panel.} 

\figcaption[fg20.eps]{The weighted mean fitting residuals of all the
primary and secondary standard stars contained within an individual CCD image
from the night ``wiync'' are plotted as a function of the Universal Time of
the observation, for the $B$, $V$, $R$, and $I$ bandpasses (top to bottom).
The number in the bottom right corner of each panel indicates the total 
number of stellar observations (in chips~1 and 2 combined) that went into
these averages.}

\figcaption[fg21.eps]{The weighted mean $V$-band fitting residuals of
all the Landolt (1992) photometric standard stars observed within an individual
CCD image has been plotted against the airmass at which the observation was
obtained, for those nights of WIYN 3.5m observations that are considered
photometric.  As in Figs.~18 and 19, small filled circles represent results
for chip~1 of MiniMos, and large empty circles are for chip~2.  Error bars
($\pm1\sigma$) are also plotted at the position of each point.  The number
in the lower right corner of each panel indicates the total number of
standard-star observations represented.  These are the data that
were employed by Saha \etal\ to estimate their extinction coefficients.  It is
evident that if only these stars are used, in some cases is is possible to infer
an extinction coefficient that is appreciably different from the one I adopted.} 

\figcaption[fg22.eps]{The same as Fig.~21, except for the $I$ photometric bandpass.}

\figcaption[fg23.eps]{The differences in the calibrated $V$-band
magnitudes obtained on the five photometric WIYN 3.5m nights relative to my
adopted standard values for primary and secondary standard stars observed with
MiniMos chip~1 have been binned in 0.5-mag intervals of \vmi\ color.  The median
residual within each bin has been plotted against the mean \vmi\ color of the
stars in the bin (filled circles), and the error bars represent a robust measure
of the dispersion among the residuals within that bin.  As in Figs.~18 through
22, the data have been divided by night:  ``wiyn2'' (top), ``wiyn3,'',
``wiyn4,'' ``wiynb,'' and ``wiync'' (bottom).  Precisely the same color
transformation was employed for all five nights, and also for the nonphotometric
night ``wiyna'' (not illustrated).  The number of different standard stars
represented is given in the lower right corner of each panel.}

\figcaption[fg24.eps]{The same as Fig.~23, except for the $V$ filter and
chip~2.}

\figcaption[fg25.eps]{The same as Figs.~23 and 24, except for the $I$ filter
and chip~1.}

\figcaption[fg26.eps]{The same as Figs.~23 through 25, except for the $I$ filter
and chip~2.}

\figcaption[fg27.eps]{The $V$-band magnitude differences between
my calibrated magnitudes and Landolt's (1992) published values for stars
measured with MiniMos chip~1 on the five photometric nights of observations
with the WIYN 3.5m telescope.  Where a given Landolt standard has been
observed more than once with chip~1 on a given night, I have used the best
weighted average of my calibrated magnitudes.  These are the data that
were employed by Saha \etal\ to derive their color transformations.  It is
clear that in a few cases, if these were the only data used, it would be
possible to estimate zero-points and color transformations perceptibly different
from the ones I adopted.  The number of different Landolt standards represented
is given in the lower right corner of each panel.}

\figcaption[fg28.eps]{The same as Fig.~27, except for the $V$ filter and chip~2.}

\figcaption[fg29.eps]{The same as Figs.~27 and 28, except for the $I$ filter and chip~1.}

\figcaption[fg30.eps]{The same as Figs.~27 through 29, except for the $I$ filter and chip~2.}

\clearpage

\baselineskip 0.1 true cm
\footnotesize

\begin{deluxetable}{llllrrrrrr}
\tablecaption{Photometric datasets for \ngc{2419}}
\tablecolumns{10}
\tablewidth{0pt}
\tablehead{
\colhead{Observing run} & \colhead{Telescope} & \colhead{Detector} &
\colhead{Year/Month} & \colhead{Clr} &
\colhead{Cld} & \colhead{$B$} & \colhead{$V$} & \colhead{$R$} & \colhead{$I$}
}
\startdata
nbs          &  CTIO 4m   & RCA1    & 1983 Jan      &  5 &  -- & 14 & 14 &  3 &  2 \\
jvw          &  INT 2.5m  & RCA     & 1986 Mar/Apr  & -- &   1 &  2 &  2 &  2 &  2 \\
cmr          &  INT 2.5m  & GEC     & 1986 Oct      & -- &   1 &  5 &  5 &  4 &  6 \\
igs          &  INT 2.5m  & GEC4    & 1989 Mar/Apr  & -- &   1 &  1 &  1 &  2 &  1 \\
psb          &  INT 2.5m  & EEV5    & 1992 Mar      &  3 &   1 &  4 &  3 &  6 &  1 \\
rjt          &  INT 2.5m  & EEV5    & 1992 Apr/May  & -- &   1 &  6 &  6 &  6 &  3 \\
brewer       &  CFHT 3.6m & lick2   & 1991 Sep      & -- &   1 & -- &  4 & -- &  2 \\
bond         &  KPNO 4m   & t2kb    & 1994 Apr      &  1 &  -- &  4 &  4 & -- &  4 \\
bolte        &  KPNO 2.1m & t1ka    & 1994 Apr      &  1 &   1 & -- & 20 & -- & 22 \\
aaj          &  INT 2.5m  & EEV5    & 1994 Nov      &  1 &  -- & -- &  6 & -- &  8 \\
mxt          &  INT 2.5m  & TEK3    & 1995 Apr      &  2 &  -- &  2 &  2 &  2 &  1 \\
wiyna        &  WIYN 3.5m & MiniMos & 2000 Sep      & -- &   2 & -- &  6 & -- &  6 \\
wiync        &  WIYN 3.5m & MiniMos & 2001 Feb      &  2 &  -- &  5 &  6 &  5 &  6 \\
wiynb        &  WIYN 3.5m & MiniMos & 2001 Apr      &  2 &  -- & -- & -- & -- & -- \\
wiyn4        &  WIYN 3.5m & MiniMos & 2001 Sep      &  2 &  -- & -- &  4 & -- &  4 \\
wiyn3        &  WIYN 3.5m & MiniMos & 2001 Dec      &  2 &  -- & -- &  3 & -- &  5 \\
abi36        &  WIYN 0.9m & s2kb    & 2002 Nov      &  3 &  -- &  3 &  3 &  3 &  3 \\
wiyn2        &  WIYN 3.5m & MiniMos & 2003 Feb      &  2 &  -- &  4 &  2 &  2 &  2 \\
\enddata
\end{deluxetable}
%
%
%
%
%
%
%
%
%
%
%
%
%
%
%
%
%
%
%
%
%
%
%
%
%
%
%
%
%
%
%
%
\end{document}